\documentclass[journal, one column, 12pt]{IEEEtran}
\usepackage{mathpazo}
\usepackage{times}
\usepackage{amsmath}
\usepackage{amsfonts}
\usepackage{latexsym}
\usepackage{amssymb}
\usepackage{mathabx}
\usepackage{float}
\usepackage{upref}
\usepackage{theorem}
\usepackage{graphicx}
\usepackage{subcaption}
\usepackage{psfrag}
\usepackage{cite}
\usepackage{cancel}

\usepackage{color}
\usepackage{tikz}
\usetikzlibrary{arrows.meta,bending}
\makeatletter
\newcommand\curvearrowed@[3]
  {%
    \begin{tikzpicture}
      \node[inner sep=.05ex](a){\kern-.25ex#3\kern-.05ex};
      \draw[arrows={-Latex[#1]},line width=#2]
        (a.north west) to[bend left=45] (a.north east);
    \end{tikzpicture}%
  }
\newcommand\curvearrowed[1]
  {%
    \relax
    \ifmmode
      \mathchoice
        {\curvearrowed@{}{.4pt}{$\displaystyle #1$}}
        {\curvearrowed@{}{.4pt}{$\textstyle #1$}}
        {\curvearrowed@{scale=.8}{.325pt}{$\scriptstyle #1$}}
        {\curvearrowed@{scale=.7}{.25pt}{$\scriptscriptstyle #1$}}%
    \else
      \curvearrowed@{}{.4pt}{#1}%
    \fi
  }
\makeatother

\usepackage{algorithm,algpseudocode}

\makeatletter
\newcommand{\removelatexerror}{\let\@latex@error\@gobble}
\makeatletter
\newcommand{\proofpart}[2]{%
	\par
	\addvspace{\medskipamount}%
	\noindent\emph{Part #1: #2}\par\nobreak
	\addvspace{\smallskipamount}%
	\@afterheading
}
\makeatother

\theoremstyle{plain}
\theorembodyfont{\normalfont\slshape}

\newtheorem{thm}{Theorem$\!$}
\newenvironment{theorem}
{\begin{thm}\hspace*{-1ex}{\bf.}}{\end{thm}}

\newtheorem{clm}[thm]{Claim$\!$}
\newenvironment{claim}{\begin{clm}\hspace*{-1ex}{\bf.}}{\end{clm}}

\newtheorem{lem}[thm]{Lemma$\!$}
\newenvironment{lemma}{\begin{lem}\hspace*{-1ex}{\bf.}}{\end{lem}}

\newtheorem{prop}[thm]{Proposition$\!$}

\newtheorem{cor}[thm]{Corollary$\!$}
\newenvironment{corollary}{\begin{cor}\hspace*{-1ex}{\bf.}}{\end{cor}}

\newtheorem{defn}[thm]{Definition$\!$}

\newtheorem{xmpl}[thm]{Example$\!$}
\newenvironment{example}{\begin{xmpl}\hspace*{-1ex}{\bf.}}{\hfill $\Box$ \end{xmpl}}

\newtheorem{cnstr}{Construction$\!$}

\newtheorem{rmk}[thm]{Remark$\!$}

\newcounter{enumrom}
\renewcommand{\theenumrom}{(\roman{enumrom})}

\makeatletter
\renewcommand{\@endtheorem}{\endtrivlist}
\makeatother

\makeatletter
\renewcommand{\thefigure}{{\@arabic\c@figure}}
\renewcommand{\fnum@figure}{{\bf Figure\,\thefigure}}
\makeatother

\newcommand{\cA}{\mathcal{A}}
\newcommand{\cB}{\mathcal{B}}
\newcommand{\cC}{\mathcal{C}}
\newcommand{\cD}{\mathcal{D}}

\newcommand{\cM}{\mathcal{M}}

\newcommand{\cO}{\mathcal{O}}

\newcommand{\cS}{\mathcal{S}}
\newcommand{\cT}{\mathcal{T}}

\newcommand{\ba}{\mathbf{a}}

\newcommand{\bb}{\mathbf{b}}

\newcommand{\bd}{\mathbf{d}}

\newcommand{\br}{\mathbf{r}}
\newcommand{\bs}{\mathbf{s}}
\newcommand{\bt}{\mathbf{t}}
\newcommand{\bu}{\mathbf{u}}
\newcommand{\bv}{\mathbf{v}}

\newcommand{\bz}{\mathbf{z}}

\newcommand{\bfa}{{\boldsymbol a}}

\newcommand{\bfc}{{\boldsymbol c}}
\newcommand{\bfd}{{\boldsymbol d}}

\newcommand{\bfs}{{\boldsymbol s}}

\newcommand{\bfx}{{\boldsymbol x}}
\newcommand{\bfy}{{\boldsymbol y}}
\newcommand{\bfz}{{\boldsymbol z}}

\renewcommand{\leq}{\leqslant}

\renewcommand{\geq}{\geqslant}

\newcommand{\Cref}[1]{Co\-ro\-lla\-ry\,\ref{#1}}

\ifCLASSINFOpdf
\else
\fi

\begin{document}
%
\title{Reconstruction of Sets of Strings from Prefix/Suffix Compositions}

\author{Ryan~Gabrys,~\IEEEmembership{Member,~IEEE,}
        Srilakshmi~Pattabiraman,~\IEEEmembership{Student~Member,~
        IEEE}
        and~Olgica~Milenkovic,~\IEEEmembership{Fellow,~IEEE}
\thanks{R. Gabrys, S. Pattabiraman and O. Milenkovic are with the Department
of Electrical and Computer Engineering, University of Illinois, Urbana-Champaign, Urbana, IL, USA. e-mail: ryan.gabrys@gmail.com, sp16@illinois.edu, milenkov@illinois.edu.}
\thanks{Parts of the results were presented at Information Theory Workshop (ITW) 2020, Riva del Garda, Italy.}}

%
%
\markboth{Journal of \LaTeX\ Class Files,~Vol.~13, No.~9, September~2014}%
{Shell \MakeLowercase{\textit{et al.}}: Bare Demo of IEEEtran.cls for Journals}
%
\maketitle

\begin{abstract}
The problem of reconstructing strings from substring information has found many applications due to its importance in genomic data sequencing and DNA- and polymer-based data storage. One practically important and challenging paradigm requires reconstructing mixtures of strings based on the union of compositions of their prefixes and suffixes, generated by mass spectrometry devices. We describe new coding methods that allow for unique joint reconstruction of subsets of strings selected from a code and provide upper and lower bounds on the asymptotic rate of the underlying codebooks. Our code constructions combine properties of binary $B_h$ and Dyck strings and that can be extended to accommodate missing substrings in the pool. As auxiliary results, we obtain the first known bounds on binary $B_h$ sequences for arbitrary even parameters $h$, and also describe various error models inherent to mass spectrometry analysis. This paper contains a correction of the prior work by the authors, published in \cite{gabrys2020mass}. In particular, the bounds on the prefix codes are now corrected.
\end{abstract}


%
\IEEEpeerreviewmaketitle

\section{Introduction}

\IEEEPARstart{M}{odern} digital data storage systems are facing fundamental storage density limits and to 
address the emerging needs for large volume archiving, it is of importance to identify new nanoscale recording media. Recently proposed DNA-based data storage paradigms~\cite{al2017mass,goldman2013towards,grass2015robust,yazdi2017portable, yazdi2015rewritable,launay2020precise,tabatabaei2020dna,pan2021rewritable} offer storage densities that are orders of magnitude higher than those of flash and optical recorders but the systems often come with a prohibitively high cost and slow and error-prone read/write platforms. To mitigate the issues associated with potentially ambiguous data reconstruction and to correct a diverse type of errors inherent to DNA sequencing technologies, several new coding solutions that aid in string assembly, dealing with asymmetries in the readout channel, and reconciliation of multiple string evidence sets were introduced in~\cite{kiah2016codes,gabrys2017asymmetric,gabrys2017codes,gabrys2019unique,gabrys2020set,agarwal2020group,cheraghchi2020coded} (see also the related and follow-up lines of work~\cite{jain2017duplication,raviv2018rank,abroshan2019coding,lenz2019anchor,chang2017rates,shomorony2021dna}). 

As an alternative to DNA-based data storage systems, polymer-based data storage systems~\cite{al2017mass,launay2020precise} are particularly attractive due to their low cost~\cite{al2017mass}. In such platforms, two molecules of significantly different masses are synthesized to represent the bits $0$ and $1$, respectively. The molecules are used as building blocks in the sequential process of recording user-defined information content. The obtained synthetic polymers are read by tandem mass (MS/MS) spectrometers. A mass spectrometer breaks multiple copies of the polymer uniformly at random, thereby creating prefixes and suffixes of the string of various lengths. The readout system outputs masses of these prefixes and suffixes. 
If the masses of all prefixes from a single string are accounted for and error-free, reconstruction is straightforward. But if multiple strings are read simultaneously and the masses of prefixes and suffixes of the same length are confusable, the problem becomes significantly more complicated. It is currently not known which combinations of coded binary strings can be distinguished from each other based on prefix-sufix masses and for which code rates is it possible to perform unique multistring reconstruction.

In a related research direction, the problem of reconstructing a string from an abstraction of its MS/MS output was considered in~\cite{acharya2014string}, under the name \textit{string reconstruction from its substring composition multiset}. The \textit{composition} of a binary string is the number of $0$s and the number of $1$s in the string. For example, the composition of $001$ equals $0^2 1^1$, indicating that $001$ contains two $0$s and one $1$, without revealing the order of the bits. The substring composition multiset $C(\bs)$ of a string $\bs$ is the multiset of compositions of all possible substrings of the string $\bs$. As an illustration, the set of all substrings of $001$ equals $\{ 0,0,1,00, 01, 001\}$, and the substring composition multiset of $001$ equals $\{ 0^1,0^1,1^1,0^2, 0^11^1, 0^21^1\}$. Two modeling assumptions are used for the purpose of rigorous mathematical analysis of this problem~\cite{acharya2014string} and in subsequent works~\cite{pattabiraman2019reconstruction,gabrys2020mass, pattabiraman2020coding}: a) Using MS/MS measurements, one can uniquely infer the composition of a polymer substring from its mass; and b) When a polymer is broken down for mass spectrometry analysis, the masses of all its substrings are observed with identical frequencies. 

Under the above modeling assumptions, the authors of~\cite{acharya2014string} established that strings are uniquely reconstructable up to reversal provided that the length of the strings $n$ is one less than a prime or one less than twice a prime, or whenever $n\leq 7$. The work~\cite{pattabiraman2019reconstruction,gabrys2020mass,pattabiraman2020coding} demonstrated that at most logarithmic code redundancy can ensure unique reconstruction of single strings drawn from codebooks based on Bertrand-Catalan strings or Reed-Solomon-like constructions.

However, the assumption that MS/MS output measurements include masses of all substrings is often not true in practice, as breaking the string in one rather than two locations is easier to perform. In the former case, one is presented with masses of the prefixes and suffixes. Thus, for the string $001$, one would observe the multiset $\{ 0^1,\cancel{0^1},1^1,0^2, 0^11^1, 0^21^1\}$. Furthermore, in practice the content of multiple strings are often read simultaneously, which complicates the matter even further as it is not known a priori which prefixes and suffixes are associated with a given string.

The problem addressed in this work may be formally stated as follows. We seek the size of the largest code $C(h)$ of binary strings of a fixed length $n$ with a property we refer to as \emph{$h$-unique reconstructability}. For any subcollection $\bs_1, \bs_2, \ldots,\bs_{\bar{h}}$ of $\bar{h} \leq h$ strings from $C(h)$, one is presented with the union $\cM(\bs_1) \cup \cM(\bs_2) \cup \cdots \cup \cM(\bs_{\bar{h}})$ of the prefix-suffix composition multisets, $\cM(\bs_i), \, i=1,\ldots,\bar{h},$ of the individual strings $\bs_i, \, i=1,\ldots,h$. The prefix-suffix composition multiset $\cM(\bs)$ of a string $\bs$ captures the weights of prefixes and suffixes of the string $\bs$ of all lengths. Unique reconstruction refers to the property of being able to distinguish all possible $h'$-unions and nonambiguously determine the identity of the strings in the collection. Our main result provides a construction for $C(h)$ that asymptotically approaches $1/h$, under certain mild parameter constraints. The proofs of our results rely on the use of Dyck and binary $B_h$ strings. For the latter, only constructions and bounds pertaining to $h=2$ and $h=n$ have been known in the literature~\cite{lindstrom1975determining,lindstrom1969determination,lindstrom1972b2}, while we provide new results for arbitrary even values of $h$. We also introduce a simple scheme for combating missing prefix-suffix errors in the pool and motivate the study of a number of new error-control coding problems associated with mixture reconstructions.

The paper is organized as follows. Section~\ref{sec:intro} introduces the problem, as well as the relevant terminology and notation. Section~\ref{sec:lbmc} describes the code constructions and the corresponding lower-bound analysis for the code rate. Upper bounds are presented in Section~\ref{sec:ubmc}. Error-control coding schemes are described in Section~\ref{sec:ecc}, along with open problems.

\section{Problem Statement and Preliminaries}\label{sec:intro}

We start by introducing the relevant notation. Let $\bs= s_1 \ldots s_n \in \{0,1\}^n$ be a binary string of length $n$ and let  $\cM(\bs)$ denote the composition multiset of all prefixes and suffixes of $\bs$. For example, if $\bs = 01101$, then 
\begin{align*}
    \cM(\bs) = \Big \{ 0, 01, 01^2, 0^21^2, 0^21^3, 1, 01, 01^2, 01^3, 0^21^3 \Big \}.
\end{align*}
We denote the set of prefix and suffix compositions of $\bs$ as $\cM_p(\bs)$ and $\cM_s(\bs)$, respectively. For the above string, $\cM_p(\bs) = \{ 0, 01, 01^2, 0^21^2, 0^21^3 \}$ and $\cM_s(\bs) = \{ 1, 01, 01^2, 01^3, 0^21^3  \}$.

We seek to design a binary codebook $C(n,h) \subseteq \{0,1\}^n$ so that for any collection of distinct strings $\bs_1, \bs_2, \ldots, \bs_{\bar{h}}  \in C(n,h)$ with ${\bar{h}} \leq h,$ the multiset 
\begin{align}\label{eq:constraint}
    \cM(\bs_1) \cup \cM(\bs_2) \cup \cdots \cup \cM(\bs_{\bar{h}})
\end{align}
uniquely determines the individual strings in the collection. We refer to a code that satisfies~(\ref{eq:constraint}) as an \textit{\textbf{$h$-multicomposition code},} or an \textit{\textbf{$h$-MC code}}. For simplicity, we often use $\cM(S)$ to describe the multi-composition set for $S = \{\bs_1, \bs_2, \ldots, \bs_h\}$. We also say that $C_p (n,h) \subseteq \{0,1\}^n$ is an $h$-prefix code if for any two distinct sets of size $\leq h$, say $S_1, S_2 \subseteq C_p$, 
$ \cM_p(S_1) \neq \cM_p(S_2).$ 

The next claim establishes a useful connection between our problem and the related problem of determining binary strings based on their real-valued sum. 
\begin{claim}\label{cl:sumsets} 
Given $\cM_p(\bs_1) \cup \cM_p(\bs_2) \cup \cdots \cup \cM_p(\bs_h)$, one can determine the real-valued sum $\bs_1 + \bs_2 + \cdots + \bs_h$.
\end{claim} 
\begin{IEEEproof} We prove the result for $h=2$ as the generalization is straightforward. Suppose that $\bs_1, \bs_2 \in \{0,1\}^n$. Then, given $\cM_p(\bs_1) \cup \cM_p(\bs_2)$, let $n_i$ denote the total number of ones in the two compositions of prefixes of length $i$ in the multiset (i.e., sum of their weights). It is straightforward to see that $\bs_1 + \bs_2  = t_1 t_2 \ldots t_n$, where $t_i = n_i - n_{i-1}$, with $n_0=0$.    
\end{IEEEproof}

\begin{example} \label{ex:sum-recovery}
As an illustrative example, consider the strings $\bs_1=110100$ and $\bs_2=101010$, for 
which we have $\bs_1+\bs_2=211110$. Clearly, $(\bs_1+\bs_2)_1=2$, which we obtained by summing up the compositions of prefixes of length one, i.e., $1+1=2$. Next, it is easy to see that $(\bs_1+\bs_2)_2=(\bs_1+\bs_2)_1^2-(\bs_1+\bs_2)_1$, where for simplicity of notation we used $(\bs_1+\bs_2)_1^2$ to denote the sum of the weights of the prefixes of length two.  A straightforward calculation reveals that $(\bs_1+\bs_2)_2=(2+1)-2=1$. The other positions in the sum of the strings can be determined similarly. 
\end{example}

The above claim provides a useful connection between our problem and the problem of designing binary $B_h$ sequences. A \emph{binary $B_h$ sequence} is a set $\cS_h(n)$ of binary strings of fixed length $n$ such that for any two distinct subsets of strings in $\cS_h(n)$, say $\cS = \{ \bs_1, \bs_2 , \ldots, \bs_{{\bar{h}}_1}\} \neq \cT=\{\bt_1, \bt_2, \ldots, \bt_{{\bar{h}}_2}\},$ where ${\bar{h}}_1, {\bar{h}}_2 \leq h$, one has
\begin{align}\label{eq:b2seq}
\sum_{i=1}^{{\bar{h}}_1} \bs_{i} \neq \sum_{j=1}^{{\bar{h}}_2} \bt_j.
\end{align}
Here, addition is performed over the reals. To avoid possible confusion with the naming convention involving sequences of sequences, we henceforth refer to the above entity as collection of binary $B_h$ sequences or binary $B_h$ codes of length $n$.

\begin{example} \label{eg:bh}
Consider the set $\cS_2(6) =\{ 110100,101010,110010 \}$. It is easy to verify that the real-valued sums of pairs of strings in $\cS_2(6)$ are distinct. Thus, $\cS_2(6)$ is a binary $B_2$ code.

However, $\cS'_2(6) = \{ 110100,101010,110010, 101100\}$ is not a binary $B_2$ code since $110100+101010 = 110010+101100 = 211110.$
\end{example}
Based on Claim~\ref{cl:sumsets}, it is easy to identify two sufficient conditions for a set of strings to be an $h$-MC code: 
\begin{enumerate}
\item \textbf{Condition 1:} One can recover $\cM_p(\bs_1) \cup \cdots \cup \cM_p(\bs_h)$ from $\cM(\bs_1) \cup \cdots \cup \cM(\bs_h)$, for any choice of $h$ distinct codestrings $\bs_1, \ldots, \bs_h$; and 
\item \textbf{Condition 2:} The codestrings $\bs_1, \ldots, \bs_h$ belong to a binary $B_h$ code $\cS_h(n)$. 
\end{enumerate}
These observations will be used to construct $h$-MC codes in Section~\ref{sec:lbmc}. Note that the condition that the codestrings in an MC code belong to a $B_h$-code is not necessary. For example, consider the case $\bs_1 = 011$, $\bs_2 = 000,$ $\bs_3 = 001$, $\bs_4 = 010$. Then,
$ \bs_1 + \bs_2 = 011 = \bs_3 + \bs_4,$
but $01^2 \in \cM(\bs_1) \cup \cM(\bs_2)$ and $01^2 \not \in \cM(\bs_3) \cup \cM(\bs_4)$, so that $\{\bs_1, \bs_2\}$ and $\{\bs_3, \bs_4\}$ are not confusable. However, a direct consequence of Claim~\ref{cl:sumsets} is that the maximum size of a $B_h$ code is at most the maximum size of a $h$-prefix code.

The best currently known upper bound on the rate of binary $B_2$ codes is $.5753$~\cite{cohen2001binary}. For $h>2$ such that $h \neq n$, we are unaware of any other bounds on the rate of binary $B_h$ codes other the ones presented in this work. 
We show next that for sufficiently large code lengths, the maximum rate of an $h$-MC code is at least $\frac{1}{h}$. 

\section{A Constructive Lower Bound for $h$-MC Codes}\label{sec:lbmc}

We start with a binary $B_h$ code and introduce redundancy into the underlying strings to ensure that given the multi-composition set of at most $h$ strings, one can separate the prefixes from the suffixes. Then, given the set of prefixes, one can use the same idea in Claim~\ref{cl:sumsets} to recover the sum of the $h$ codestrings and hence the codestrings themselves.

Let $\cS_h(n) \subseteq \mathbb{F}_2^n$ be a $B_h$ code over $\mathbb{F}_2^n$. It is well-known that $\cS_h(n)$ can be constructed using the columns of a parity-check matrix of a code with minimum Hamming distance $\geq 2h+1$. Using this construction, we have 
$$ \lim_{n \to \infty} \frac{1}{n} \log |\cS_h(n)| = \frac{1}{h}.$$

For our problem and the underlying approach for solving it, we will also make use of Dyck strings: A string $\bs  \in \mathbb{F}_2^N$ of even length $N$ is a 
Dyck string if its weight satisfies $\text{wt}(\bs) =  \frac{N}{2},$ and for $i \in [N-1]$,
\begin{align}\label{eq:dyck1}
\text{wt}(s_1 s_2\ldots s_i) \geq \Big \lceil \frac{i}{2} \Big \rceil.
\end{align}

The approach for generating the code $C(N,h) \subseteq \mathbb{F}_2^{N}$ is to ensure that it satisfies the following two properties:
\begin{enumerate}
\item A string $\bs \in C(N,h)$ is a Dyck string;
\item The set $C(N,h)$ is a binary $B_h$ code of length $N$.
\end{enumerate}
The first property ensures that the mixtures of prefixes and suffixes can be partitioned into two sets, one containing all the prefixes and another containing all the suffixes. The second property ensures that given the prefix set (or, alternatively, the suffix set) once can recover the codestrings using the simple observation that the prefixes uniquely determine the real-valued sum of the strings in the mixture. We illustrate these observations with an example. 

\begin{example} \label{eg:sum} 
Consider the binary $B_2$ code $\cS_2(6) =\{ 110100,101010,110010 \}$. Clearly, all three strings are Dyck strings as their prefixes of any length contain at least as many ones as zeros.

Next, write $\bs_1 = 110100$ and $\bs_2 = 101010$, so that $\cM(\bs_1) \cup \cM(\bs_2) =$ $\{1,1, 01,$ $1^2, 01^2,$ $01^2, 0^2 1^2,$ $01^3, 0^21^3, 0^21^3,$ $0^31^3, 0^31^3,$ $0^31^3, 0^31^3, $ $0^31^2, 0^31^2,0^21^2, 0^31,  0^21, 0^21, 01, 0^2, 0, 0 \}$. Since $ \bs_1$ and $\bs_2$ are Dyck strings, each of the string prefixes must have at least as many $1$s as $0$s. Similarly, each suffix must have at least as many $0$s as $1$s. It follows from this observation that one can easily recover the multiset $\cM_p(\bs_1) \cup \cM_p(\bs_2)  = $ $\{ 1,1, 01,$ $1^2, 01^2,$ $01^2, 0^2 1^2,$ $01^3, 0^21^3, 0^21^3,$ $0^31^3, 0^31^3 \}.$

Claim~\ref{cl:sumsets} ensures that given $\cM_p(\bs_1) \cup \cM_p(\bs_2),$ one can determine $\bs_1 + \bs_2 = 211110$. Since $\cS_2(6)$ is a binary $B_2$ code, the sum $\bs_1 + \bs_2$ uniquely determines the strings $\bs_1$ and  $\bs_2$.
\end{example}

The next claim establishes the formal result that if the code $C(N,h)$ satisfies these two properties, then it is an $h$-MC code.

\begin{claim}\label{cl:dyckrec} Suppose that $C(N,h)$ is a $B_h$ code where for any $\bs \in C(N,h)$, Equation (\ref{eq:dyck1}) holds. Then, $C(N,h)$ is an $h$-MC code.
\end{claim}
\begin{IEEEproof} Similar to Claim~\ref{cl:sumsets}, we prove the statement for the case where $h=2$, since the extension for general $h$ is straightforward. In light of Claim~\ref{cl:sumsets}, we need to show that the property in (\ref{eq:dyck1}) allows us to uniquely recover $\cM_p(\bs_1) \cup \cM_p(\bs_2)$ from $\cM(\bs_1) \cup \cM(\bs_2)$. To see that this is indeed possible, observe that from (\ref{eq:dyck1}) both prefixes of length $i$ in $\cM(\bs_1) \cup \cM(\bs_2)$ have at least $\lceil \frac{i}{2} \rceil$ $1$s whereas both suffixes of length $i$ in $\cM(\bs_1) \cup \cM(\bs_2)$ have at most $\lfloor \frac{i}{2} \rfloor$ $1$s. 
\end{IEEEproof}

To construct codes $C(N,h)$ of large cardinality, we perform a simple ``balancing procedure'' on each codesting $\bs \in \cS_h(n)$ and then append $\cO(\sqrt{n})$ bits of redundancy to the beginning and end of $\bs$ so that the resulting string has length $N = n + \cO(\sqrt{n})$. Note that under this setup, it follows that for any $\epsilon$, we have
$$ \frac{1}{N} \log |C(N,h)| = \frac{1}{n + \kappa \sqrt{n}} \log |\cS_h(n)| =  \frac{1}{h} - \epsilon,$$
where $\kappa$ is a constant, and $\epsilon>0$ can be made arbitrarily small for $n$ sufficiently large.

To maximize the rate of the coding scheme and combine the two constraints that $h$-\textbf{MC} strings need to satisfy, we use two ideas. First, we use $B_h$ strings parsed into blocks that allow us to tightly control the weights of the codestrings. Second, rather than working directly with the weights of strings as described in (\ref{eq:dyck1}), we use the running digital sums (RDSs). For a string $\bs \in \mathbb{F}_2^n$, the RDS up to coordinate $i$ is defined as $R(\bs)_i = 2\text{wt}(s_1 s_2 \ldots s_i) - i$. If the subscript $i$ is omitted, then $R(\bs) = 2\text{wt}(\bs) - |\bs|$, where $|\bs|$ denotes the length of $\bs$. Using the running digital sum, the constraint in Equation (\ref{eq:dyck1}) can be rewritten as $\text{wt}(\bs) = \lceil \frac{N}{2} \rceil$ and $R(\bs)_i \geq 0,$ $i \in [N]$.

\begin{figure}[h!]
  \centering
  \begin{subfigure}[b]{0.8\linewidth}
    \includegraphics[width=\linewidth]{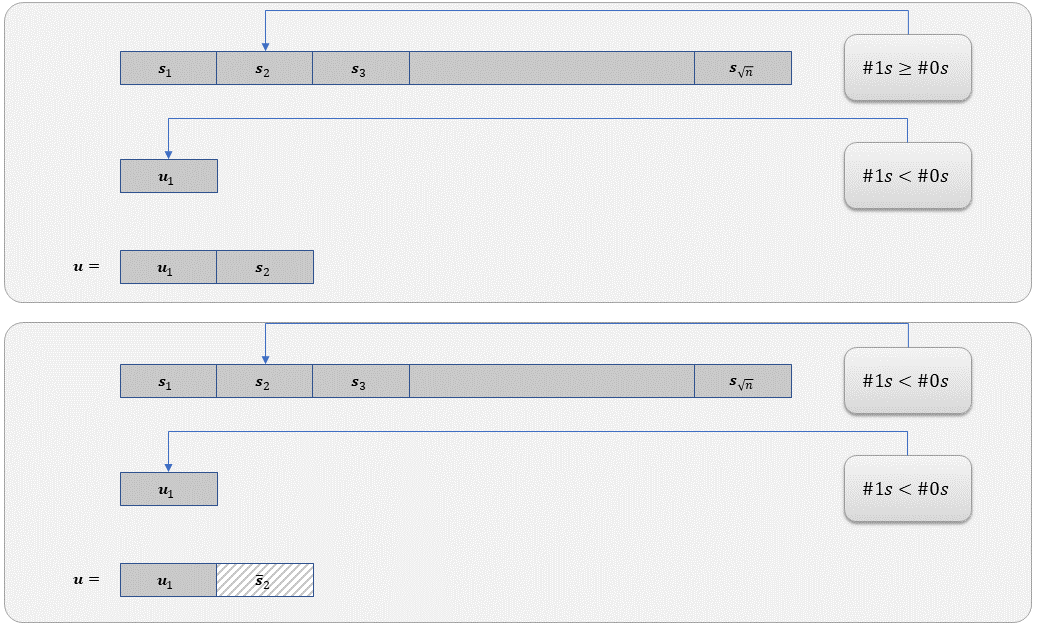}
    \caption{In the first step, we set $\bu_1 = \bs_1$. Without loss of generality, let us assume that $\bu_1$ contains more $0$s than $1$s. Then, the next block $\bu_2$ is set to $\bs_2$ if $\bs_2$ has more $1$s than $0$s. Otherwise, $\bu_2$ is set to the complement of $\bs_2$, $\bar{\bs}_2$. To reconstruct a block $\bu_2,$ we either use the complement of $\bs_2$ or the string itself, as determined by the added flag bit $r_2$.} 
  \end{subfigure} \hfill
  \begin{subfigure}[b]{0.8\linewidth}
    \includegraphics[width=\linewidth]{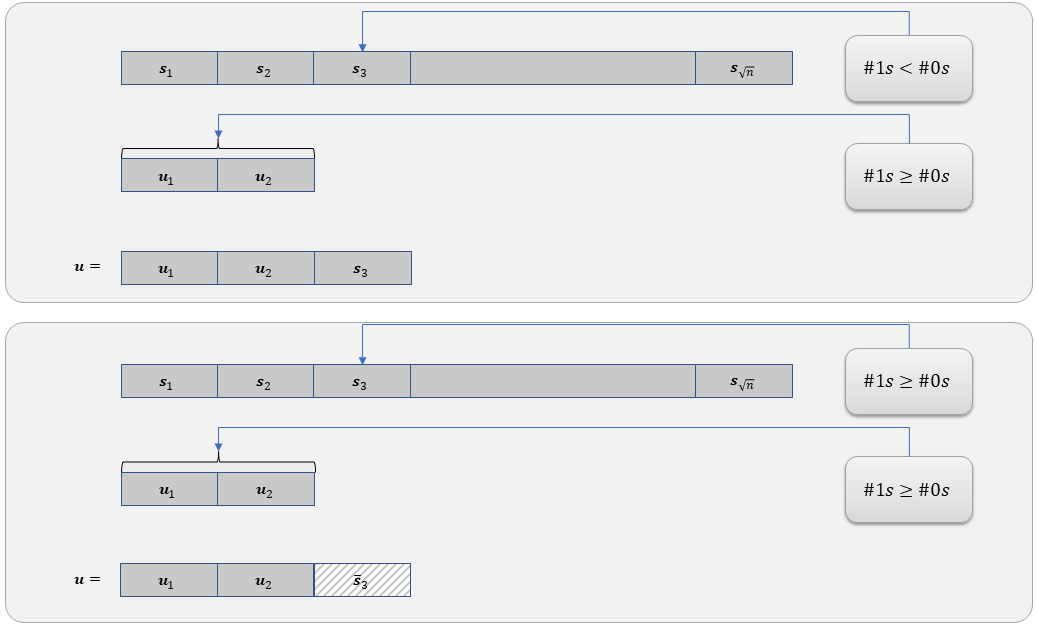} 
    \caption{In the next step, the approximate balancing procedure continues as follows: Let us assume that $\bu_1 \bu_2$ has more $1$s than $0$s. The next block $\bu_3$ equals $\bs_3$ if $\bs_3$ has more $0$s than $1$s. Otherwise, 
    $\bu_3$ equals the complement of $\bs_3$, $\bar{\bs}_2$. The procedure is repeated for all subsequent steps.}
\end{subfigure}
\caption{Illustration of the approximate balancing procedure. The string $\bs = \bs_1 \bs_2 \dots \bs_{\sqrt{n}}$ is decomposed into a concatenation of $\sqrt{n}$ blocks, each of length $\sqrt{n}$. The string $\bu = \bu_1 \bu_2 \dots \bu_{\sqrt{n}}$ is constructed sequentially from the blocks $\bs_1, \bs_2,  \dots , \bs_{\sqrt{n}}$ by ensuring that the discrepancy between the number of $1$s and $0$s in the reconstructed string $\bu_1 \bu_2 \dots \bu_i$ is reduced or kept the same at every step $i$. Subfigure (a) depicts how the block $\bu_2$ is chosen, while subfigure (b) depicts how the block $\bu_3$ is chosen.}
  \label{fig:1}
\end{figure}

\begin{figure}
  \includegraphics[width=\linewidth]{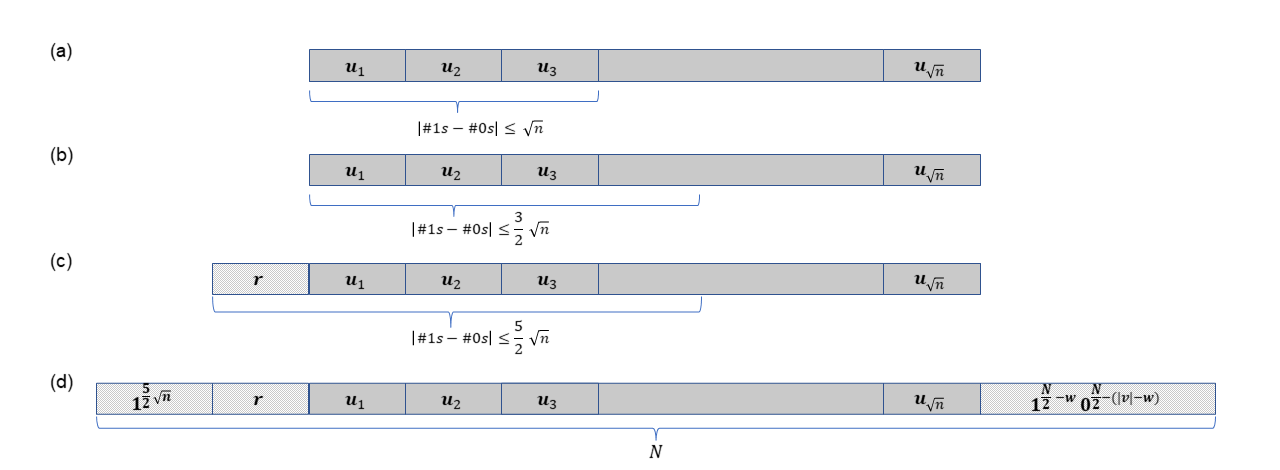}
  \caption{Further examples regarding the code construction. Subfigure (a) illustrates the finding of Claim~\ref{cl:bbal}: The discrepancy between the number of $1$s and $0$s in the string $\bu_1 \dots \bu_i, $ $i \in [\sqrt{n}]$ is at most $\sqrt{n}$. Subfigure (b) illustrates the finding of Lemma~\ref{lem:rdsz}: The discrepancy between the number of $1$s and $0$s in the string $u_1 u_2 \dots u_j ,$ $j \in [n]$ is at most $\frac{3}{2}\sqrt{n}.$ Subfigure (c) illustrates an immediate extension of Lemma~\ref{lem:rdsz}: The discrepancy between the number of $1$s and $0$s in the string $\br u_1 u_2 \dots u_j ,$ $j \in [n]$ is at most $\frac{5}{2}\sqrt{n}.$ Recall that $\bv = {\bf1}^{5/2 \sqrt{n}} \br \bu $ is as defined in Claim~\ref{cl:bal2}, and that  $w = \text{wt}(\bv)$. Subfigure (d) depicts how adding a properly chosen prefix and suffix to the string ensures the Dyck property. } 
  \label{fig:4}
\end{figure}

The balancing procedure operates as follows: Let $\bs \in \cS_h(n)$, and for simplicity, assume that $\sqrt{n}$ is an even integer. We begin by parsing $\bs$ into blocks $\bs_i$ of length $\sqrt{n},$ $i=1,\ldots, \sqrt{n}$, so that $\bs = \bs_1 \bs_2 \ldots \bs_{\sqrt{n}}  \in \mathbb{F}_2^n$. Using $\bs$ we construct an auxiliary string $\bu=\bu_1 \bu_2 \ldots \bu_{\sqrt{n}}$ that is ``approximately'' balanced following an idea similar to Knuth's balancing, which operates on blocks rather than individual symbols (please refer to Figures~\ref{fig:1} and~\ref{fig:4} for an illustration). We start by initializing $\bu_1 = \bs_1$. For a binary string $\bu$, we use $\overline{\bu}$ to denote the binary complement of $\bu$. For $j \in \{2,3,\ldots,\sqrt{n}\}$, we define $\bu_j$ according to:
\begin{align}\label{eq:procdbal}
\bu_j = \begin{cases}
\bs_j,& \text{ if } R(\bu_1 \ldots \bu_{j-1}) < 0, \text{ and } R(\bs_j) \geq 0,\\
\overline{\bs}_j,& \text{ if } R(\bu_1 \ldots \bu_{j-1}) < 0, \text{ and } R(\bs_j) < 0,\\
\bs_j,& \text{ if } R(\bu_1 \ldots \bu_{j-1}) \geq 0, \text{ and } R(\bs_j) < 0,\\
\overline{\bs}_j,& \text{ if } R(\bu_1 \ldots \bu_{j-1}) \geq 0, \text{ and } R(\bs_j) \geq 0.
\end{cases}
\end{align}

The next claim immediately follows from~(\ref{eq:procdbal}).

\begin{claim}\label{cl:bbal} For any $j \in [\sqrt{n}]$, $|R(\bu_1 \ldots \bu_j)| \leq \sqrt{n}.$ Hence, the RDS of complete collections of subblocks is bounded in absolute value by $\sqrt{n}$.
\end{claim}

We also have the following result.
\begin{lemma}\label{lem:rdsz} For any $i \in [n]$, $ |R(\bu)_i| \leq \frac{3}{2} \sqrt{n}.$ Hence, the RDS of any prefix of $\bu$ does not exceed $\frac{3}{2} \sqrt{n}$ in absolute value.
\end{lemma}
\begin{IEEEproof} Suppose, on the contrary, that $|R(\bu)_i| > \frac{3\sqrt{n}}{2}$. For simplicity, we will only consider the case $R(\bu)_i > \frac{3\sqrt{n}}{2}$, as the other case can be handled similarly. Next, assume that $j \in [n]$ is the smallest index for which $R(\bu)_j > \frac{3\sqrt{n}}{2}$ and that we may assume that $R(\bu)_j = \frac{3\sqrt{n}}{2} +1$. Now, let $j = k_1 \sqrt{n} + k_2,$ where $0 \leq k_2 < \sqrt{n}$. According to Claim~\ref{cl:bbal}, since $R(\bu)_j = \frac{3\sqrt{n}}{2}+1$ and $k_2 < \sqrt{n}$, we have 
\begin{align}\label{eq:RDS1}
\frac{\sqrt{n}}{2}< R(\bu_1 \ldots \bu_{k_1-1}) \leq \sqrt{n}.
\end{align}
Based on (\ref{eq:procdbal}), and since $R(\bu_1 \ldots \bu_{k_1-1}) > \frac{\sqrt{n}}{2}$,  it follows that $R(\bu_{k_1}) \leq 0$ so that 
\begin{align}\label{eq:RDS2}
- \sqrt{n}  < R(\bu_{k_1})_{\ell} \leq \frac{\sqrt{n}}{2}
\end{align}
for any $\ell \in [\sqrt{n}]$. Combining (\ref{eq:RDS1}) and (\ref{eq:RDS2}), we have that $R(\bu)_{k_1 \sqrt{n} + k_2} \leq \frac{3 \sqrt{n}}{2}$, which is a contradiction.
\end{IEEEproof}

We now describe our encoder. Let $\bu \in \{0,1\}^n$ be the string which is the result of the procedure described in (\ref{eq:procdbal}), and suppose that $\br \in \{0,1\}^{\sqrt{n}}$ is such that for any $j \in [\sqrt{n}]$:
\begin{align}\label{eq:uzlabel}
\br_j = \begin{cases}
1, & \text{ if } \bu_j \neq \bs_j,\\
0, & \text{ if } \bu_j = \bs_j.
\end{cases}
\end{align}
Using $\br$, we now form a string $\bs \in C(N,h)$, where $N = n + \frac{17}{2} \sqrt{n} $, and assume for 
simplicity that $N$ is an even integer. The following claim is used in our subsequent analysis.

\begin{claim}\label{cl:bal2} Let $\bv = {\bf1}^{5/2 \sqrt{n}} \br \bu \in \{0,1\}^{n + 7/2 \sqrt{n} }$. Then, for any $i \in [n + \frac{7}{2} \sqrt{n} ]$,
$$ |R(\bv)_i| \leq 5 \sqrt{n} .$$
Furthermore, for any $i \in [n + 7/2 \sqrt{n}]$, 
$$ R(\bv)_i > 0. $$
\end{claim}

We now append redundant bits to the string $\bv$ described in Claim~\ref{cl:bal2} in order to get a string $\bs \in \{0,1\}^{N}$ which is a Dyck string\footnote{Splitting the string into blocks of length $m$ and then performing the ``approximate'' balancing task over these blocks would incur a redundancy of $\frac{n}{m} + c m$, where $c$ is a constant. The redundancy is minimized when the summands are of the same order, $\sqrt{n}$. This justifies the use of our partition choice.}. This results in the following claim.

\begin{claim}\label{cl:ydef} Let $N =n + \frac{17}{2} \sqrt{n} $ be an even integer and let $\bv = {\bf1}^{5/2 \sqrt{n}} \br \bu \in \{0,1\}^{n + 7/2 \sqrt{n}}$ be as defined in Claim~\ref{cl:bal2}. Suppose that $w = \text{wt}(\bv)$. Then, the string 
$ \bs =  \bv {\bf1}^{\frac{N}{2} - w} {\bf0}^{\frac{N}{2}-(|\bv| - w)}  $
is a Dyck string.
\end{claim}

Now, assume that $C(N,h) \subseteq \mathbb{F}_2^N$ is constructed according to the procedure outlined in Claim~\ref{cl:ydef} and once again assume that $N = n + \frac{17}{2} \sqrt{n} $ is an even integer. The next theorem is the main result of this section and it establishes the correctness of our construction through the description of a simple decoding algorithm.

\begin{theorem}\label{thm:prf} Suppose that $\bs_1, \bs_2, \ldots, \bs_h \in C(N,h)$, where $C(N,h)$ is constructed according to the balancing procedure operating on $B_h$ strings. Then, given $ \cM(\bs_1) \cup \cM(\bs_2) \cup \cdots \cup \cM(\bs_h)$, we can uniquely determine $\{ \bs_1, \ldots, \bs_h \}$. Furthermore, for any $\epsilon >0$, there exists $n_\epsilon >0$ such that for all $N \geq n_\epsilon,$ $ \frac{1}{N} \log |C(N,h)| \geq  \frac{1}{h} - \epsilon.$
\end{theorem}

\begin{IEEEproof} For simplicity, we prove the result for $h=2$, as the extension for general values $h$ is straightforward. According to Claims~\ref{cl:dyckrec} and~\ref{cl:ydef}, we can recover $\cM_p(\bs_1) \cup \cM_p(\bs_2)$ from $ \cM(\bs_1) \cup \cM(\bs_2)$ since $\bs_1, \bs_2$ are Dyck strings. From $\cM_p(\bs_1) \cup \cM_p(\bs_2)$, we can recover $\bs_1 + \bs_2$ according to Claim~\ref{cl:sumsets}. Given $\bs_1 =  {\bf1}^{5/2 \sqrt{n} } \br_1 \bu_1 {\bf1}^{\frac{N}{2} - w_1}  {\bf0}^{\frac{N}{2} - (|\bv_1| - w_1)}$ and $\bs_2 =  {\bf1}^{5/2 \sqrt{n} } \br_2 \bu_2  {\bf1}^{\frac{N}{2} - w_2}  {\bf0}^{\frac{N}{2} - (|\bv_2| - w_2)}$, from the first $n + \frac{7}{2} \sqrt{n} $ coordinates of $\bs_1 + \bs_2$ we can recover
\begin{align*}
\left( \br_1 + \br_2, \bu_1 + \bu_2 \right) \bmod 2.
\end{align*}
Next, for shorthand, write $\bu = \bu_1 + \bu_2 \bmod 2 = \bu_1 \bu_2 \ldots \bu_{\sqrt{n}}$ and $\br = \br_1 + \br_2 \bmod 2 = r_1 \ldots  r_{\sqrt{n}}$. Let $\tilde{\bu} = \tilde{\bu}_1 \ldots \tilde{\bu}_{\sqrt{n}}$. Then, for $j \in [\sqrt{n}]$, 
\begin{align*}
\tilde{\bu}_j = \begin{cases}
\bu_j, & \text{ if } r_j = 0,\\
\overline{\bu}_j & \text{ if } r_j = 1.
\end{cases}
\end{align*}
It is straightforward to verify from~(\ref{eq:procdbal}) that $\tilde{\bu} = \bs_1 + \bs_2 \bmod 2$. Since $\bs_1, \bs_2 \in \cS_2(n)$ are $B_2$ strings over $\mathbb{F}_2^n$, we can recover $\bs_1$ and $\bs_2$ from $\tilde{\bu}$, which concludes the proof.
\end{IEEEproof}

Note that our lower bound from Theorem~\ref{thm:prf} relies on the use of binary $B_h$ codes of rate $\frac{1}{h}$. It is currently not known whether this bound is achievable and the only known results in this area pertain to the case $h=2$ and $h=n$. In the next subsection, to better our understanding of the maximum asymptotic rate of $h$-MC code, we derive new upper bounds for $B_h$ sequences, with $2<h<n$, that significantly outperform naive counting arguments. Note that these results imply upper bounds on the rates achievable using the construction from Theorem~\ref{thm:prf}. For example, it is well known that the maximum rate of a binary $B_2$ sequence is at most $0.5753$ which implies that the maximum rate of any code produced using the constructive procedure in the previous section is at most $0.5753$.

\subsection{New Upper Bounds on Binary $B_h$ Sequences}

We extend next the idea used in~\cite{lindstrom1972b2} to obtain an upper bound on the maximum rate of a binary $B_h$ code (as already pointed out, we are unaware of any results pertaining to the rate of binary $B_h$ codes for the case $h>2$ and $h \neq n$). Hence, the results contained in this section may be of independent interest. In addition, the derived bounds represent an upper bound on the rate of codes constructed in this section. Throughout this part of the text, we assume that $h$ is even. We first introduce relevant notation before describing our main result in Theorem~\ref{eq:UBBh}.

Let $B_h(n)$ denote a binary $B_h$ sequence of length $n$. Since $B_h(n)$ is a binary $B_h$ sequence it follows that for any two distinct sets of codewords, say $\{\bs_1, \bs_2, \ldots, \bs_h \}, \{\bs'_1, \bs'_2, \ldots, \bs'_h \} \subseteq B_h(n)$, we have
$\sum_{j=1}^h \bs_j \neq \sum_{j=1}^h \bs'_j,$ where addition is over the reals. As will be discussed in more details below, rather than work directly with the prefixes of codewords as suggested in~\cite{lindstrom1972b2} for the case $h=2$, we instead use the sum of prefixes of codewords in a binary $B_h$ code. Then, through information-theoretic arguments (which avoid the more complicated convolution-based approach from~\cite{lindstrom1972b2}), we arrive at Theorem~\ref{eq:UBBh}. 

Let $\cA_{h/2}$ denote the set of sums of any collection of $h/2$ prefixes of distinct codewords from $B_h(n),$ where $n = a+b$ and $a$ and $b$ are the lengths of the prefixes and suffixes, respectively. More specifically, let
\begin{align*}
 \cA_{h/2} = \left \{ \ba_1 + \ba_2 + \cdots + \ba_{h/2} \in \{0,1,\ldots,h/2\}^{a} :   (\ba_1 \bb_1), (\ba_2, \bb_2), \ldots, (\ba_{h/2}, \bb_{h/2})  \in  B_h(n)  \right \}.
\end{align*}
For $\bfc \in \cA_{h/2}$, define $\cB_{\bfc}$ as
\begin{align*}
\cB_{\bfc} = \left \{ \bb_1 + \cdots + \bb_{h/2} \in \{0,1,\ldots,h/2\}^{b} :  (\ba_1 \bb_1), \ldots, (\ba_{h/2}, \bb_{h/2})  \in B_h(n), \ba_1 + \cdots \ba_{h/2} = \bfc \right \}.
\end{align*}

We start by proving the next claim.

\begin{claim}\label{cl:diffs} Let $\bfc_1, \bfc_2 \in A_{h/2}$. Then for any $\bfd_1, \bfd_2 \in \cB_{\bfc_1}$ and $\bfd_3, \bfd_4 \in \cB_{\bfc_2}$,
\begin{align*}
\bfd_1 - \bfd_2 \neq \bfd_3 - \bfd_4,
\end{align*}
where the subtraction is performed over the reals unless $\bfd_1 = \bfd_2$ and $\bfd_3 = \bfd_4$.
\end{claim}
\begin{IEEEproof} Suppose, on the contrary, that the previous equation holds with equality. Then $ (\bfc_1 \bfd_1) + (\bfc_2 \bfd_4) = (\bfc_2 \bfd_3) + (\bfc_1 \bfd_2)$. In this case, we may write
\begin{align*}
\ba^{(1)}_1 + \ba^{(1)}_2 + \cdots + \ba^{(1)}_{h/2} &= \bfc_1,\ \ba^{(2)}_1 + \ba^{(2)}_2 + \cdots + \ba^{(2)}_{h/2} = \bfc_2, \\
 \bb^{(1)}_1 + \bb^{(1)}_2 + \cdots + \bb^{(1)}_{h/2} &= \bd_1,\ \bb^{(2)}_1 + \bb^{(2)}_2 + \cdots + \bb^{(2)}_{h/2} = \bfd_2, \\
 \bb^{(3)}_1 + \bb^{(3)}_2 + \cdots + \bb^{(3)}_{h/2} &= \bfd_3,\ \ \bb^{(4)}_1 + \bb^{(4)}_2 + \cdots + \bb^{(4)}_{h/2} = \bfd_4.
\end{align*}
Then, if the statement in the claim holds, it follows that
\begin{align*}
\left( (\ba^{(1)}_1 \bb^{(1)}_1) + \cdots + (\ba^{(1)}_{h/2} \bb^{(1)}_{h/2}) \right)& + \left( (\ba^{(2)}_1 \bb^{(4)}_1) + \cdots + (\ba^{(2)}_{h/2} \bb^{(4)}_{h/2}) \right) = \\
&\left( (\ba^{(2)}_1 \bb^{(3)}_1) + \cdots + (\ba^{(2)}_{h/2} \bb^{(3)}_{h/2}) \right) + \left( (\ba^{(1)}_1 \bb^{(2)}_1) + \cdots + (\ba^{(1)}_{h/2} \bb^{(2)}_{h/2}) \right),
\end{align*}
where $(\ba^{(1)}_1 \bb^{(1)}_1), \ldots, (\ba^{(1)}_{h/2} \bb^{(1)}_{h/2}) , \ldots, (\ba^{(1)}_{h/2} \bb^{(2)}_{h/2}) \in B_h(n)$, a contradiction.
\end{IEEEproof}

Based on the result of the previous claim, we focus on the differences between elements in the set $\cB_{\bfc}$ and define the multiset
\begin{align*}
\cD = \{{ {\bfd}_1 - {\bfd}_2  \in \{-h/2, \cdots, 0,\ldots, h/2 \}^{b} : \exists {\bfc} \text{ s.t. } {\bfd}_1, {\bfd}_2 \in \cB_{\bfc}   \}}.
\end{align*}

The next claim characterizes the properties of sequences in the multiset $\cD$.

\begin{claim}\label{cl:dfreq0} Any nonzero $\bd \in \{-h/2, \ldots, h/2 \}^{b}$ appears at most once in $\cD$, while the all-zero vector appears $\binom{|B_h(n)|}{h/2}$ times.
\end{claim}
\begin{IEEEproof} The first statement follows immediately from Claim~\ref{cl:diffs}. For the second claim, since $B_h(n)$ is a $B_h$ code, it follows that the sum of any distinct $h/2$ codewords from $B_h(n)$ is necessarily unique and hence $\bfd_1 - \bfd_2 = {\bf0}$ if and only if $\bfd_1 = \bfd_2$. Since $\sum_{\bfc \in \cA_{h/2}} |B_{\bfy}| = \binom{|B_h(n)|}{h/2}$, the result follows.
\end{IEEEproof}

Let $H_n(h)$ denote 
\begin{align*}
\frac{1}{n} \log \left( \left | \left \{\bfx_1 + \bfx_2 + \cdots + \bfx_h : \bfx_1, \bfx_2, \ldots, \bfx_h \in \{0,1\}^n \right \} \right |   \right) = H_n(h).
\end{align*}

Also, let $H(h)$ denote the entropy of the binomial distribution $B(h,\frac{1}{2})$
\begin{align}\label{eq:entro}
H\left(h \right) = - \sum_{k=0}^h \binom{h}{k} \left( \frac{1}{2} \right)^{h} \log_2 \left( \binom{h}{k} \left( \frac{1}{2} \right)^h \right).
\end{align}
It is well-known that  
$$\lim_{n \to \infty} H_n(h) = H(h), $$
and that
$$ H(h) = \frac{1}{2} \log_2 \left( 2 \pi e \frac{h}{4} \right) + \cO\left( \frac{1}{h} \right).$$

The following claim will be useful in characterizing the size of $\cD$. For simplicity, define the \emph{set} 
\begin{align*}
\cD_s =  \left \{ \bd : \bd \in \cD  \right \}
\end{align*}
containing the elements in $\cD$, with repeats of the all-zero string removed. Recall that $b$ stands for the length of the strings in $\cD$.

\begin{claim} For $b$ large enough, $\frac{1}{b} \log \left( \left| \cD_s \right| \right)$ can be approximated by $H(h)$.
\end{claim}
\begin{IEEEproof} In order to bound the size of the set $\cD_s$, we consider the size of the set:
\begin{align*}
\left\{ \left( \bfx_1 + \bfx_2 + \cdots + \bfx_{h/2} \right) - \left( \bfx'_1 + \bfx'_2 + \cdots + \bfx'_{h/2} \right) : \bfx_1, \bfx_2, \ldots, \bfx_{h/2}, \bfx'_1, \bfx'_2, \ldots, \bfx'_{h/2} \in \{0,1\}^b   \right \},
\end{align*}
where addition is over the reals. Let 
\begin{align*}
\bfz = \left(z_1, z_2, \ldots, z_b \right) = \left( \bfx_1 + \bfx_2 + \cdots + \bfx_{h/2} \right) - \left( \bfx'_1 + \bfx'_2 + \cdots + \bfx'_{h/2} \right).
\end{align*}
where $ \bfx_1, \bfx_2, \ldots, \bfx_{h/2}, \bfx'_1, \bfx'_2, \ldots, \bfx'_{h/2}$ are randomly chosen strings in $\{0,1\}^b$. Then, for $j \in \{-h/2, -h/2+1, \ldots, h/2\}$, 
\begin{align*}
\text{Pr}\left( z_i = j \right) &= \sum_{\ell=j}^{h/2} \binom{h/2}{\ell} \binom{h/2}{\ell-j} \left( \frac{1}{2} \right)^{h} = \binom{h}{h/2 + j} \left( \frac{1}{2} \right)^{h}. 
\end{align*}
The entropy of this distribution is
\begin{align*}
- \sum_{k=0}^h \binom{h}{k} \left( \frac{1}{2} \right)^{h} \log_2 \left( \binom{h}{k} \left( \frac{1}{2} \right)^h \right),
\end{align*}
which according to (\ref{eq:entro}) equals $H(h)$. This completes the proof.
\end{IEEEproof}

\begin{theorem}\label{eq:UBBh} For $h$ even, the maximum asymptotic rate of a binary $B_h$ code is at most
$$\frac{\frac{2}{h} H(h/2)}{1 + \frac{H(h/2)}{H(h)}}.$$
\end{theorem}
\begin{IEEEproof} By definition, 
\begin{align*}
\binom{\left | B_h(n) \right|}{h/2} = |\cB_{{\bfy}_1}| + |\cB_{{\bfy}_2}| + \cdots + | \cB_{{\bfy}_{|\cA_{h/2}|}}|.
\end{align*}
From the previous equation and the convexity of the function $x^2$, we have
\begin{align*}
\left| \cD \right| = \sum_{j=1}^{|\cA_{h/2}|} | \cB_{{\bfy}_j}|^2 \geq \frac{\binom{|B_h(n)|}{h/2}^2}{|\cA_{h/2}|} \approx \frac{ \left| B_h(n) \right|^h}{|\cA_{h/2}|}.
\end{align*}
Then, since according to Claim~\ref{cl:dfreq0} the all-zero vector appears at most $|B_h(n)|^{h/2}$ times and 
all the other vectors appear at most once in $\cD$,
\begin{align*}
\left| B_h(n) \right|^h &\lessapprox \left| \cD \right | \left| \cA_{h/2} \right| \\
&= |\cD \setminus \cD_s| |\cA_{h/2}| + |\cD_s| |\cA_{h/2}|\\
&\leq |B_h(n)|^{h/2} 2^{a H\left(h/2 \right) } + 2^{b H(h) + a H(h/2) }.
\end{align*}
 
Letting $|B_h(n)|^{h/2} = 2^{a H(h/2)}$, and setting $b = a \frac{H(h/2)}{H(h)}$, we obtain $| B_h(n)  |^{h/2} \lessapprox 2^{a H(h/2)} + 2^{b H(h)}$, which implies
\begin{align*}
\log \left| B_h(n) \right| \lessapprox \frac{2}{h} a H(h/2).
\end{align*}
Therefore,
\begin{align*}
\frac{\log{\left| B_h(n) \right |}}{n} \lessapprox \frac{\frac{2a}{h} H(h/2) }{a + b} = \frac{\frac{2}{h} H(h/2)}{1 + \frac{H(h/2)}{H(h)}}.
\end{align*}
\end{IEEEproof}

Clearly, a naive upper bound on the maximum size of a binary $B_h$ code is $\frac{1}{h} H(h)$, and it is straightforward to verify that for $h$ large enough the expression in the previous theorem coincides with this bound. However, for finite $h$, the previous bound is significantly tighter. For example, for $h=4,6,8,$ $\frac{1}{h} H(h)$ is approximately $.5118, .3899$, and $.3184$, respectively, whereas the bound derived above gives $.4406, .3433$ and $.2837$, respectively. 

\section{Upper Bounds on $h$-MC Codes}\label{sec:ubmc}

Next, we derive an upper bound on the maximum rate of an $h$-MC code. To this end, recall that $C_p \subseteq \{0,1\}^n$ is an $h$-prefix code if for any two subsets of sizes ${\bar{h}}\leq h$, say $\cS_1, \cS_2 \subseteq C_p$,  $ \cM_p(\cS_1) \neq \cM_p(\cS_2).$  Let $C^{(MC)}_h(n)$ be the size of the largest $h$-MC code of codelength $n$ and suppose that $C^{(p)}_h(n)$ is the size of the largest $h$-prefix code of codelength $n$. Formally, we use $R_h^{(MC)}$ to denote the maximum asymptotic rate of an $h$-MC code,
$$R_h^{(MC)} =  \lim_{n \to \infty} \sup \frac{1}{n} \log | C^{(MC)}_h(n) |.$$

We show next that when $h$ is even, $R^{(MC)}_{h} \leq 1- \frac{1}{2} \left( \frac{1}{1 + \frac{1}{h}} \right)$. Once again, for simplicity of exposition, we focus on the case $h=2$ before considering the general result.

The next lemma states that in order to derive an upper bound on the quantity $R_h^{(MC)}$, we can limit our attention to prefix codes. The result follows since the set of all suffixes is a function of the set of all prefixes provided the total number of ones in each codeword is the same and known beforehand. 

\begin{lemma}\label{lem:pset} For any $\epsilon >0$, there exists an $n_\epsilon > 0$ such that for all $n \geq n_\epsilon$, one has
$$\frac{1}{n} \log | C_h^{(MC)}(n) | \leq  \frac{1}{n} \log | C_h^{(p)}(n) | + \epsilon.$$
\end{lemma}
\begin{IEEEproof} To simplify the discussion, we focus on the case where $h=2$; the extension to $h > 2$ are straightforward. For $w \in [n]$, let $C_2^{(w)}(n) \subseteq C_h^{(MC)}(n)$ denote the set of codewords of weight $w$ in $C_2^{(MC)}(n)$. By the pigeon-hole principle, there exists a $w^{*} \in [n]$ where $|C_2^{(w^{*})}(n)| \geq \frac{1}{n} |C_2^{(MC)}(n)|$. Given two codewords in $C_2^{(w^{*})}(n)$, say $\cS =\{\bs_1, \bs_2\}$, we can easily determine $\cM_p(\cS)$. Assuming that only the prefix composition set is known, the set $\cM_s(\cS)$ can be derived as follows. To determine the compositions of suffixes of length $i$, for $i \in [n]$, we subtract from $w^{*}$ the number of ones in each prefix of length $n-i$. For instance, suppose the compositions of prefixes of length $n-1$ of $\bs_1, \bs_2$ are $\Big \{ \{ 1^{w^*}, 0^{n-{w^*}} \}, \{ 1^{w{^*}-1}, 0^{n-{w^*}+1} \} \Big \}.$ Then, the length-$1$ suffixes of $\bs_1, \bs_2$ are $\Big \{ \{1\}, \{0\} \Big \}$. This implies that $n|C_2^{(p)}(n)| \geq n| C_2^{(w^*)}(n) | \geq |C_2^{(MC)}(n)|$, which establishes the desired result.
\end{IEEEproof}

Let us first turn our attention to $h=2$. For any $\bs \in C_{2}^{(p)}(n)$, we write $\bs$ as $\bs =  \ba \bb  \in C_{2}^{(p)}(n),$
where $\ba \in \{0,1\}^{\alpha n}$ equals the first $\alpha n$ symbols of $\bs$ and $\bb$ equals the last $(1-\alpha)n$ symbols of $\bs$. We represent the codewords in the codebook using a bipartite graph $G = (V_P,V_S,E)$ with 
\begin{align}\label{eq:VP}
V_P = \left \{ \ba \in \{0,1\}^{\alpha n} : \exists \bs \in C_{2}^{(p)}(n) \text{ s.t. } \bfs=\ba \bb  \right \},
\end{align} 
and 
\begin{align}\label{eq:VS}
V_S = \left \{ \bb \in \{0,1\}^{(1-\alpha) n}  : \exists \bs \in C_{2}^{(p)}(n)  \text{ s.t. } \bfs=\ba \bb  \right \}.
\end{align} 
In this setting, an edge $(v_1, v_2) \in E$, with $v_1 \in V_P$ and $v_2 \in V_S$, connects an admissible prefix (vertex in $V_P$) to an admissible suffix (vertex in $V_S$) so that every edge  corresponds to a codeword in $C_2^{(p)}$ and vice versa.

Let $w \in \{{0,1,\ldots, \alpha n\}}=[[\alpha n+1]]$. We also find it useful to work with another bipartite graph $G^{(w)} = (V^{(w)}_P, V^{(w)}_S, E^{(w)})$ whose edges are a subset of the edges in $E$. The partition of the vertices $V^{(w)} = (V^{(w)}_P, V^{(w)}_S)$ is such that $v_1 \in V^{(w)}_P$ if and only if the prefix $\bfa \in \{0,1\}^{\alpha n}$ represented by the vertex $v_1$ in $G$ has weight $w$, and in addition, $v_2 \in V^{(w)}_S$ if and only if there exists a $v_1 \in V^{(w)}_P$ such that $(v_1, v_2) \in E$. The set $E^{(w)} \subseteq E$ is such that $(v_1, v_2) \in E^{(w)}$ if $v_1 \in V^{(w)}_P$ and $v_2 \in V^{(w)}_S$.

The next result will be used in the proof of Theorem~\ref{th:ub2}.

\begin{lemma}\label{lem:conf2} The graph $G^{(w)}$ cannot contain a cycle of length four. 
\end{lemma}
\begin{IEEEproof} Suppose, on the contrary that $G^{(w)}$ contains a $4$-cycle, say 
\begin{align*}
( {\ba}_1{\bb}_1, {\ba}_2 {\bb}_2, {\ba}_1 {\bb}_2, {\ba}_2 {\bb}_1 ).
\end{align*}
Then, 
\begin{align*}
\cM_p({\ba_1} {\bb_1}) \cup \cM_p({\ba_2} {\bb_2}) = \cM_p({\ba_2} {\bb_1}) \cup \cM_p({\ba_1} {\bb_2}).
\end{align*}
To verify the above claim, note that all prefixes of length $\alpha n$ have to be the same since 
$\cM_p(\bfa_1) \cup \cM_p(\bfa_2) = \cM_p(\bfa_2) \cup  \cM_p(\bfa_1)$. Furthermore, since $\text{wt}(\ba_1) = \text{wt}(\ba_2)$ it is straightforward to verify that all prefixes of length longer than $\alpha n$ are the same amongst $\cM_p({\ba_1} {\bb_1}) \cup \cM_p({\ba_2} {\bb_2}), \cM_p({\ba_2} {\bb_1}) \cup \cM_p({\ba_1} {\bb_2})$. But this contradicts the fact that the prefixes and suffixes involved correspond to a $2$-prefix code. This establishes the claimed result.
\end{IEEEproof}

%
We are now ready to prove our upper bound on $h$-prefix codes for $h=2$. 

\begin{theorem}\label{th:ub2}  For any $\epsilon > 0$, there exists an $n_\epsilon > 0$ such that for all $n \geq n_\epsilon$ one has $\frac{1}{n} \log |C_2^{(p)}(n)| \leq \frac{2}{3} + \epsilon$. 
\end{theorem}
\begin{IEEEproof} In order to bound the number of codewords in $C_2^{(p)}(n)$, we will upper bound the number of edges in the graph $G=(V_P,V_S,E)$. To this end, we consider the maximum number of edges in the graph $G^{(w)}=(V_P^{(w)}, V_S^{(w)},E^{(w)})$. It follows from the pigeonhole principle that there exists a $w^{*} \in [[\alpha n+1]]$ such that
\begin{align*}
\left|E^{(w^{(*)})} \right| \geq \frac{\left| E \right|}{\alpha n + 1}.
\end{align*}
Thus, $\frac{1}{n} \log \left|E^{(w^{(*)})} \right|$ can be approximated by $\frac{1}{n} \log \left| C_2^{(p)}(n) \right|$ for $n$ sufficiently large.

According to the previous lemma, $G^{(w^{(*)})}$ cannot contain a $4$-cycle.
It is well-known that the number of edges in an $m_1 \times m_2$ bipartite graph without cycles of length $4$ is at most \cite{NV05}
\begin{align}\label{eq:b4cycles}
m_1 m_2^{\frac{1}{2}} + m_1  + m_2.
\end{align}
Letting $\alpha n = \frac{n}{3}$ in (\ref{eq:b4cycles}) so that $m_1 = 2^{n/3}$ and $m_2 = 2^{2n/3}$ gives 
\begin{align*}
\frac{1}{n} \log \left|E^{(w^{(*)})} \right| \leq \frac{2}{3} + \cO\left(\frac{1}{n}\right).
\end{align*}
This implies the desired result.
\end{IEEEproof}

The next corollary follows from the previous theorem and Lemma~\ref{lem:pset}.

\begin{corollary}\label{cor:2bound} A $2$-prefix code must have a rate bounded as $R^{(MC)}_2 \leq \frac{2}{3}.$
\end{corollary}

Next, we consider the extension to the case where $h>2$ based on the same approach. Let $C_{h}^{(p)}(n)$ denote an $h$-prefix code of length $n$. As before, we represent our codewords using a graph $G^{(h)}=(V_P^{(h)},V_S^{(h)},E^{(h)})$ as defined in (\ref{eq:VP}) and (\ref{eq:VS}), except that $(\ba, \bb) \in E^{(h)}$ if and only if $(\ba, \bb) \in C_{h}^{(p)}(n)$.  As before, we will also work with the bipartite graph $G^{(w,h)} = (V_P^{(w,h)}, V_S^{(w,h)}, E^{(w,h)}) \subseteq G^{(h)}$, which is restricted to only use prefixes of weight $w$. Our next lemma is a natural generalization of Lemma~\ref{lem:conf2}.

\begin{lemma}\label{lem:conf2} The graph $G^{(w,h)}$ cannot contain a $2h$-cycle. 
\end{lemma}
\begin{IEEEproof} Suppose, on the contrary, that the statement in the lemma does not hold and that $(\ba_1, \bb_1)$, $(\bb_1, \ba_2)$, $(\ba_2, \bb_2)$, $\ldots,$ $(\ba_h, \bb_h)$, $(\bb_h, \ba_1)$. Then, we have:
\begin{align}\label{eq:b1}
\ba_1 \bb_1, \ba_2 \bb_2, \ba_3 \bb_3, \ldots, \ba_h \bb_h \in C^{(p)}_h(n),
\end{align}
but also that 
\begin{align}\label{eq:b2}
\bb_1 \ba_2, \bb_2 \ba_3, \bb_4 \ba_5, \ldots, \bb_h \ba_1 \in C^{(p)}_h(n).
\end{align}
Since all the prefixes in $G^{(w,h)}$ have weight $w$, it is straightforward to verify that the set of codewords from (\ref{eq:b1}) and the set in (\ref{eq:b2}) have the same prefix composition multisets.
\end{IEEEproof}

The next result follows from the same arguments used in Theorem~\ref{th:ub2} and Corollary~\ref{cor:2bound}.

\begin{theorem}\label{th:UBh} For odd $h$, $R^{(MC)}_{h} \leq \frac{h+1}{2h}$. For even $h$, $R^{(MC)}_{h} \leq 1- \frac{1}{2} \left( \frac{1}{1 + \frac{1}{h}} \right)$.
\end{theorem}
\begin{IEEEproof} The results follows using the same arguments as those described in Theorem~\ref{th:ub2} and by noting that the maximum number of edges in a $m_1 \times m_2$ bipartite graph that does not contain a cycle of length $2h$ is at most $\left( m_1 m_2 \right)^{h+1}{h} + m_1 + m_2$ when $h$ is odd~\cite{NV05}. For the case when $h$ is even, the maximum number of edges is $m_1^{\frac{k+2}{2k}} m_2^{\frac{1}{2}} + m_1 + m_2$~\cite{NV05}.
\end{IEEEproof}
We can see that the gap between our upper bound stated in the previous theorem and the best possible rate code achievable from the previous section (using a binary $B_2$ sequence of rate at most $0.6$) is at least $.09$ when $h=2$. For larger $h$, this gap grows and for $h=4,6,8$, the gap is $.0882$, $.1815$, and $.2372,$ respectively. 

For large $h$, the bound from Theorem~\ref{th:UBh} implies that the maximum rate for any $h$-MC code is at most $\frac{1}{2}$ whereas our lower bound converges to zero. However, for small $h$, the gap between the upper and lower bounds is significantly smaller. For example, when $h=2$ and $h=3$, our upper bound equals $\frac{2}{3},$ whereas the lower bounds from the previous section equal $\frac{1}{2}$ and $\frac{1}{3}$, respectively. For $h=4$, the upper bound equals $\frac{3}{5}$ whereas the lower bound equals $\frac{1}{4}$. We would like to point out that the result in our preprint and the short conference version of the paper contains an error on page 4~\cite{GPM20} and that the claim that the lower and upper bound are asymptotically the same is incorrect. 

As a final note, we observe that the work in~\cite{lindstrom1969determination,lindstrom1972b2} also considered the case of nonbinary $B_h$ codes for $h=2$. The main result is that for a large enough alphabet, the maximum asymptotic rate of nonbinary $B_2$ codes is at most $\frac{1}{2}$. 

\section{Error Models and Error-Correction} \label{sec:ecc}

The MS/MS readout technique is error prone. Often, not all the masses of prefixes and suffixes are measured and/or reported. Furthermore, polymer fragmentation causes the loss of some atoms and create errors in the actual mass values; as a result, some fragments can gain or loose in mass value based on how the fragment was created. 

Mass errors can lead to issues in the process of mixture reconstruction and hence, in what follows, we first describe common error patterns in MS/MS readouts and describe simple techniques that can be used to correct them. In all the described error-modes, as throughout the whole text, we tacitly assume that one can actually determine the length of the prefixes/suffixes based on their masses. This is possible if the masses of $0$s and $1$s differ significantly (for example, if the masses of the $1$ or $0$ molecules differ by at least $n$) or if other design criteria are met. 

Converting practical mass errors into abstract error models is rather challenging. This is why we first describe how to model such errors, and in particular, errors that cannot be automatically detected and corrected. The focus of the first part is missing prefixes or suffixes. We start with the description of the effect of such errors on the process of reconstructing \emph{a single string}, and then proceed to describe how missing string errors affect the reconstruction of \emph{a mixture of multiple strings}. In both settings, we 
break up the discussion into two cases: One, in which we explain how to use the natural redundancy ensured by the presence of both prefixes and suffixes of the string to identify and correct errors; and another one, where we explain how to add controlled redundancy to mitigate the effect of MS/MS errors. In the latter case, we describe several simple error-control schemes that either add redundancy to the $B_h$ strings themselves, and/or use unequal error-protection for various substrings in the Dyck-$B_h$ codestrings as well. Furthermore, based on the approach that converts prefix/suffix masses into sums of codestrings, we introduce a straightforward idea for encoding that uses integrals and derivatives of strings.

Consider a prefix-suffix composition multiset $\cM(\bs)$ of a Dyck string $\bs$ of length $n$. 
A single missing composition can be identified and corrected without coding redundancy: 
Since either $\cM_p(\bs)$ or $\cM_s(\bs)$ are available for reconstruction, one of these two multisets will contain no errors and can consequently be used for error-free reconstruction. The following example illustrates that it is not always possible to reconstruct $\cM(\bs)$ of a string $\bs$ from an erroneous prefix-suffix composition multiset $\tilde{\cM}(\bs)$ that contains two or more erasures if no redundancy is used.

\begin{example} \label{eg:multie}
Let $\tilde{\cM}_p(\bs)$ and $\tilde{\cM}_s(\bs)$ denote the prefix and suffix composition multisets of $\bs$ with missing fragment errors, respectively. Furthermore, let 
$\tilde{\cM}(\bs)=\tilde{\cM}_p(\bs) \cup \tilde{\cM}_s(\bs)$.

\begin{itemize}
\item  Consider the string $\bs = 111000$. Given $$\tilde{\cM} = \{ 1, 1^3, 01^3, 0^21^3, 0^31^3, 0^31^3, 0^31^2, 0^31, 0^3, 0^2, 0\},$$ one can immediately see that a prefix composition of length two is missing. Since the weight of the string is three and the suffix composition of length $6-2=4$ is $0^31,$ it is clear that the missing composition is $1^2.$

\item Consider the same string, and the erroneous prefix-suffix composition multiset  $$\tilde{\cM} = \{ 1, 1^3, 01^3, 0^21^3, 0^31^3, 0^31^3, 0^31^2, 0^3, 0^2, 0\}.$$ Clearly, the composition of a prefix of length two and a suffix of length four are missing. However, the constraints imposed by the prefix compositions $1$ and $1^3$ imply that the composition of the prefix of length two is $1^2$. Similarly, the constraints imposed by suffix compositions $0^31^2$ and $0^3$ imply that the composition of the suffix of length four is $0^31$.

\item In the third example, let the string be $\bs = 110100,$ and the erroneous prefix-suffix composition multiset 
$$\tilde{\cM}= \{ 1, 1^2, 01^2, 0^21^3, 0^31^3, 0^31^3, 0^31^2, 0^31, 0^2, 0 \}.$$ From $\tilde{\cM}_p$ one can reconstruct the partial prefix $110\varepsilon\varepsilon0$ (where `$\varepsilon$' denotes that the corresponding bit cannot be determined from the procedure described in Claim~\ref{cl:sumsets}); similarly, from $\tilde{\cM}_s$ one can reconstruct the partial suffix string $11\varepsilon\varepsilon00$. By combining the partially reconstructed strings with erasures, we can easily recover the bits in all positions except for position four, and then using a process similar to the one described in the previous examples reconstruct $\bs = 110100$ and $\cM(\bs)$.

\item Next, let 
$$\tilde{\cM} = \{ 1, 1^2,  01^3, 0^21^3, 0^31^3, 0^31^3, 0^31^2, 0^31, 0^2, 0\}.$$ 
Note that both $\cM(111000) $ and $\cM(110100)$ are consistent with the erroneous composition multiset and hence reconstruction is impossible. 
\end{itemize}
\end{example}

Based on Example~\ref{eg:multie}, a number of observations are in place (any comment pertaining to prefixes also applies to suffixes and vice versa, due to symmetry).
\begin{itemize}
\item If $\tilde{\cM}_p$  is such that prefix compositions of lengths $i, i+1, i+2, \dots, i+j-2$ 
are erased, while compositions of prefixes of length  $i-1$, $i+j-1$ and $i+j$ remain intact, then the bits $i, i+1, i+2, \dots, i+j, i+j-1$ of the prefix string cannot be inferred using the technique describe in Claim~\ref{cl:sumsets}. Such a prefix composition list is said to have a \textit{contiguous erasure burst} of length $j$, starting at index $i$.
\item If the weight of the prefix composition of length $i-1$ is $w_{i-1}$ and that of length $i+j-1$ is $w_{i+j-1}, $ then among the bits at positions $i, i+1, i+2, \dots, i+j-2,$ exactly $w_{i+j-1} - w_i$ are $1$s, while the remaining bit values are $0$s.
\item If $\cM$ is missing $t$ compositions, then either $\cM_p$ or $\cM_s$ is missing at most $\lfloor \frac{t}{2} \rfloor $ compositions.
\end{itemize}
Thus, in the presence of $t$ prefix-suffix erasures, one is always able to reconstruct the string with at most $t$ missing bits. The erasures that occur in the prefix (suffix) strings are correlated and every erasure occurs as part of a contiguous burst of length $\geq 2$. By comparing the string reconstructed using prefixes with that using suffixes, certain types of erasures can be corrected as further illustrated in Example~\ref{eg:overlap}. In summary, if at most $t_p$ prefix compositions are missing, and at most $t_s$ suffix compositions are missing, then one needs to correct not more than $2 \text{ min } \{ t_p, t_s\}$  erasures in the prefix (suffix) string, each occurring in a contiguous burst of length at least two. As will be shown Section~\ref{sec:direct}, the length-two bursts can be completely eliminated from analysis and subsequent coding approaches by resorting to the use of integrals of strings, which amounts to running sums (over the reals) of the elements of the string~\cite{gabrys2017codes}. 

\begin{example} \label{eg:overlap}
Let us revisit Example~\ref{eg:multie}. Given the following erroneous multiset of the string 
$\bs = 110100,$ 
$$\tilde{\cM} = \{ 1, 1^2, 01^2 , 0^21^3, 0^31^3, 0^31^3, 0^31^2, 0^31, 0^2, 0 \},$$ 
from $\tilde{\cM}_p$ we reconstructed $110\varepsilon\varepsilon0$, and from $\tilde{\cM}_s$ we reconstructed $11\varepsilon\varepsilon00$. Using these two strings with erasures we were able to reconstruction the original string.

Next, assume that we are instead given the multiset $\tilde{\cM} = \{ 1, 1^2,  0^21^3, $  $0^31^3, 0^31^3, 0^31^2, 0^2, 0 \}.$ From $\tilde{\cM}'_p$ we can reconstruct $11\varepsilon\varepsilon\varepsilon0$, and from $\tilde{\cM}_s$ we can reconstruct $1\varepsilon\varepsilon\varepsilon00.$ By combining the two reconstructions we can only recover $11\varepsilon\varepsilon00.$ The multiset $\tilde{\cM}$ is consistent with both $\cM(111000)$ and $\cM(110100).$
\end{example}

A simple observation for the $t$ missing prefix/suffix model is as follows. A missing prefix composition of length $i$ can be recovered from a suffix composition of length $n-i.$ Thus, the number of error patterns that can be corrected without additional coding redundancy, independent on the string being a Dyck string or not, is at least 
$$ \sum_{i =0}^{t}  { n \choose i } {n-i \choose t-i }  \geq  { n \choose \lfloor \frac{t}{2} \rfloor } {n- \lfloor \frac{t}{2} \rfloor \choose \lceil \frac{t}{2} \rceil }.$$ 
This follows from the fact that we can choose $0 \leq i \leq t$ missing masses in the prefix set, fix those $i$ masses in the suffix set as ``observed'' and then select additional $t-i$ missing masses from the remaining $n-i$ suffixes.

Based on Example~\ref{eg:overlap}, and by noting that prefixes are red from the left, while suffixes are read from the right, define 
$$\mathcal{I}_p = \{ (i_p,j_p) \text{ s.t. } \text{ the prefix string has a contiguous erasure burst of length $j_p$ starting at index $i_p$} \},$$ and 
$$\mathcal{I}_s = \{ (i_s,j_s) \text{ s.t. } \text{ the suffix string has a contiguous erasure burst of length $j_s$ starting at index $i_s$} \}.$$ 
The prefix string is said to have a \textit{contiguous erasure overlap} of length $\ell$ with the suffix string if there exists $(i_p,j_p) \in \mathcal{I}_p$ and $(i_s,j_s) \in \mathcal{I}_s$ such that $j_s , j_p \geq \ell$ \textit{and} either $ i_p \leq i_s ,  (i_p + j_p - i_s) = \ell$  or $ i_s < i_p , (i_s + j_s - i_p) = \ell.$
We say that the prefix string and the suffix string erasures \textit{ do not overlap } if for all lengths $\ell \geq 1$ the prefix string does not have a contiguous erasure overlap with that of the suffix string.

\begin{claim}
A Dyck string $\bs$ can be reconstructed from the prefix-suffix composition multiset $\tilde{\cM}(\bs)$ without using error-correction redundancy if the missing prefixes and suffixes are such that 
\begin{itemize}
\item The prefix string and the suffix string do not overlap; or, \\
\item The prefix string has exactly one contiguous erasure overlap with the suffix string, and the overlap is of length one.
\end{itemize}
\end{claim}
We next turn our attention to describing the effect of missing prefixes and suffixes on multistring reconstruction from the union of the underlying prefix-suffix composition multiset, and detection/correction strategies that may be used without resorting to controlled error-control redundancy.

As for the case of a single string, one missing prefix-suffix composition in $\cM(\bs_1) \cup \cM(\bs_2) \cdots \cup \cM(\bs_h)$ corresponding to the Dyck strings $\bs_1, \bs_2, \dots \bs_h$ can be easily corrected as either the union of prefix composition multiset or the union of suffix composition multiset is error-free. But when more than one error is present, it is not always possible to reconstruct $\cM(\bs_1) \cup \cM(\bs_2)  \cdots \cup \cM(\bs_h)$ from the union of the erroneous prefix-suffix composition multiset $\tilde{\cM}(\bs_1) \cup \tilde{\cM}(\bs_2) \cdots \cup \tilde{\cM}(\bs_h)$ as illustrated below. Here, we find the following definitions useful. The partially reconstructed prefix of the sum of strings constructed from the union of erroneous prefix composition multisets using Claim~\ref{cl:sumsets} is referred to the \textit{partial prefix-sum string} and its analogue pertaining to suffixes is referred to the \textit{partial suffix-sum string}. 

\begin{example} \label{eg:multistringerror}
\begin{itemize}
\item Let us revisit Example~\ref{ex:sum-recovery} under the assumption that we are given the erroneous multiset 
$$\tilde{\cM}(\bt_1) \cup \tilde{\cM}(\bt_2) = \{1,1, 01, 1^2, 01^2, 0^2 1^2, 01^3, 0^21^3, 0^21^3, 0^31^3, 0^31^3, 0^31^3, 0^31^3, 0^31^2, 0^31^2, 0^21^2, 0^31, 0^21, 0^21, 01, 0^2, 0, 0 \}.$$
Clearly, the composition of a prefix of length three is missing. Thus, using the union of the suffix composition multiset, $$\tilde{\cM}_s(\bt_1) \cup \tilde{\cM}_s(\bt_2) = \cM_s(\bt_1) \cup \cM_s(\bt_2) = \{0^31^3, 0^31^3, 0^31^2, 0^31^2,0^21^2, 0^31, 0^21, 0^21, 01, 0^2, 0, 0 \},$$ 
one can easily determine $\bt_1 +\bt_2 = 211110$ using the procedure outlined in Example~\ref{ex:sum-recovery}. If one were to use $\tilde{\cM}_p(\bt_1) \cup \tilde{\cM}_p(\bt_2)$ instead, while setting aside the fact that a composition of length three is missing, the same procedure applied to the third symbol of $\bt_1+\bt_2 $ would results in $\text{wt}(01^2)  - \text{wt}(01) - \text{wt}(1^2) =-1,$ which is clearly indicative of an error. Negative values in the partial prefix-sum string or suffix-sum string is a clear indication of one or more missing compositions or composition errors in general.

\item In the next example, we assume that we are given the following erroneous multiset instead
$$\tilde{\cM}(\bt_1) \cup \tilde{\cM}(\bt_2) = \{1,1, 01, 1^2, 01^2, 0^2 1^2, 01^3, 0^21^3, 0^21^3, 0^31^3, 0^31^3, 0^31^3, 0^31^3, 0^31^2, 0^21^2, 0^31, 0^21, 0^21, 01, 0^2, 0, 0 \}.$$
It is once again easy to see that a prefix of length three and a suffix of length five are missing. Using the erroneous union of the prefix composition multisets, $\tilde{\cM}_p(\bt_1) \cup \tilde{\cM}_p(\bt_2),$ one can recover the partial prefix-sum $21\varepsilon\varepsilon10$, and similarly, using the union of the suffix composition multisets, $\tilde{\cM}_s(\bt_1) \cup \tilde{\cM}_s(\bt_2),$ one can recover the partial suffix-sum $\varepsilon\varepsilon1110$. By combining the two partial sum, one can recover $\bt_1 +\bt_2 = 211110.$ Note that the third bits in the two strings equal $0$ and $1$ \textit{or} $1$ and $0$, implying that the correct prefix compositions of length three are either $\{ 01^2, 01^2\}$ or $\{ 1^3, 0^21\}$. Since  $\{ 1^3, 0^21\} \cap \tilde{\cM}_p(\bt_1) \cup \tilde{\cM}_p(\bt_2) = \phi,$ one can conclude that the missing prefix composition is $01^2.$ A similar line of reasoning may be used to recover the missing suffix composition $0^31^2.$

\item In the third scenario, we are given the following erroneous multiset 
$$\tilde{\cM}(\bt_1) \cup \tilde{\cM}(\bt_2) = \{1,1, 01, 1^2, 01^2, 0^2 1^2, 0^21^3, 0^21^3, 0^31^3, 0^31^3, 0^31^3, 0^31^3, 0^31^2, 0^21^2, 0^31, 0^21, 0^21, 01, 0^2, 0, 0 \}.$$
Here, one prefix of length three and of length four is missing, as well as a suffix of length four and another of length five. Using the erroneous union of the prefix composition multisets, $\tilde{\cM}_p(\bt_1) \cup \tilde{\cM}_p(\bt_2)$ one can recover the partial prefix-sum $21\varepsilon\varepsilon\varepsilon0$, and similarly using the union of the suffix composition multisets, $\tilde{\cM}_s(\bt_1) \cup \tilde{\cM}_s(\bt_2),$ one can recover the partial suffix-sum $\varepsilon\varepsilon\varepsilon110$. By combining the two partial strings, we obtain $\bt_1 +\bt_2 = 21\varepsilon110.$ Since $\text{wt}(\bt_1 + \bt_2) =  \text{wt}(\bt_1) + \text{wt}(\bt_2)  = 3+3 =6,$ we must be that $\bt_1 +\bt_2 = 211110$. Hence, as for the case of a single string, one can recover the sum of the mixture strings even when both the partial prefix and suffix contain errors.

\item In the last example to consider, assume that we are given the erroneous multiset 
$$\tilde{\cM}(\bt_1) \cup \tilde{\cM}(\bt_2) = \{1,1, 01, 1^2,  01^2, 0^2 1^2, 01^3, 0^21^3, 0^21^3, 0^31^3, 0^31^3, 0^31^3, 0^31^3, 0^31^2, 0^31^2, 0^21^2, 0^31, 0^21, 01, 0^2, 0, 0 \}.$$ 
In this case, the prefix and suffix of length three are missing. Using the erroneous union of the prefix composition multisets, $\tilde{\cM}_p(\bt_1) \cup \tilde{\cM}_p(\bt_2),$ one can recover the partial prefix-sum $21\varepsilon\varepsilon10$, and, similarly, using the union of the suffix composition multisets, $\tilde{\cM}_s(\bt_1) \cup \cM_s'(\bt_2),$ one can recover the partial suffix-sum $21\varepsilon\varepsilon10$. However, for this case one cannot recover the missing compositions or the sum of the two input strings: Both $\cM(111000) \cup \cM(101010)$ and $\cM(110100) \cup \cM(101010)$ are consistent with the given input erroneous composition multiset.
\end{itemize}
\end{example}
We hence have the following claim.

\begin{claim}
The sum over the reals of $h$ Dyck strings $\bs_1, \bs_2, \dots , \bs_h$ of length $n$ can be reconstructed from the union of their prefix-suffix composition multiset with erasures $\tilde{\cM}(\bs_1) \cup \tilde{\cM}(\bs_2) \cup \dots \cup \tilde{\cM}(\bs_h)$ if \\
1) The prefix sum string and the suffix sum string do not overlap; or,\\
2) The prefix sum string has exactly one contiguous erasure overlap with the suffix sum string, and the overlap is of length one.
\end{claim}

As before, we remark that the multiset of all $h$ prefix compositions of length $1 \leq i \leq n$ can be recovered provided the complete multiset of $h$ suffix compositions of length $n-i$ is available. Hence, the number of error patterns that can be corrected without additional redundancy is at least the number of error patterns such that the missing prefix compositions of length $i$ can be recovered from the complete set of suffix compositions of length $n-i$ and vice-versa. To that end,

1) Assume that erasures occur in compositions of exactly $1 \leq j \leq t$ different lengths. There are ${n \choose j} $ different ways to choose these lengths.

2) For every $i \in [n]$, either the erasures are contained in the multiset of prefix compositions of length $i$ or the multiset of suffix compositions of length $n-i$. Thus, there are $2^j$ different prefix-suffix composition error patterns for the $j$ different lengths. 

3) An erroneous prefix-suffix composition multiset comprising strings of length $i$ contains between $1$ and $h$ errors. 
Thus, the number of error patterns restricted to $j$ composition lengths equals the number of positive solutions of the equation $t = t_1 + t_2 + \dots + t_j, $ such that $t_1, t_2 ,\dots , t_j \leq h,$ and is well known to be $\binom{t-1}{j-1}$ provided that $t \leq h$.

Thus, assuming that the number of missing oligos is $t \leq h$ and by adding up all possible contributions for different choices of $j$, one can find a simple lower bound on the number of error patterns that be corrected without additional redundancy
$$\sum_{j=0}^{t} {n \choose j} \binom{t-1}{j-1}\, 2^j=\sum_{j=0}^{t} {n \choose j} \binom{t-1}{t-j}\, 2^j.$$

\subsection{One-Step Error-Correction}

In what follows, we present a simple scheme that can correct up to $t$ missing prefix-suffix composition errors. Recall that the $B_h$ codebook $\cS_h(n),$ described in Sections II and III, can be constructed using the columns of a parity-check matrix of a code with minimum Hamming distance $d \geq 2h+1$. The idea behind our error-correction technique is to ensure that the real-valued sum of every $h$-string subset of the code is an error-tolerant codestring. An approach to this problem was first proposed in~\cite{Gritsenko2015On} for the purpose of designing signature codes for the noisy MAC problem (i.e., codes capable of correcting errors in the syndrome of a received word) and it consists of encoding the columns of a parity-check matrix $\tilde{H}^{k \times n},$ capable of correcting $h$ substitution errors, using a linear binary code that can correct $\lfloor \frac{t}{2} \rfloor$ substitution errors. Note that the parameter $t$ can be chosen independently from the parameter $h$ as long as $\lfloor \frac{t}{2} \rfloor \leq k \leq n$. For encoding purposes, the authors suggest using two binary BCH codes, so that $\tilde{H}$ is the parity check matrix of a BCH code of designed distance $\geq 2h+1$, while the parity-check matrix used to introduce error-control redundancy to the columns of $\tilde{H}$ is also chosen according to a BCH code with dimension $k$, length $n$ and redundancy not exceeding $\lfloor \frac{t}{2} \rfloor \, \log(n+1),$ and capable of correcting at least $\lfloor \frac{t}{2} \rfloor$ substitution errors. Here, and elsewhere in this section all logarithms are taken base two unless stated otherwise. Clearly, the only difference is that in our setting, we only encounter erasures in the coded strings (the augmented columns), and hence can handle $t$ erasures. 

Note that this construction, as pointed out by the authors, does not fully exploit the fact that addition is performed over the reals and not over the field
$\mathbb{F}_2$. As in the previous construction, $\tilde{H}$ is the parity-check matrix of a binary $h$ substitution error-correction code. But instead of using another binary code to protect the syndromes,~\cite{Gritsenko2015On} suggests finding the smallest prime $p>h$, and using a linear code over $\mathbb{F}_p$ (e.g., a Reed-Solomon code of length $p-1$) for the syndrome error-control redundancy. The dimension of the latter code equals $n$, and it is required that the code be able to correct $t$ substitution errors over the field $\mathbb{F}_p$. Since the redundancy is nonbinary, each symbol of the parity-check string is converted into a string of length $\log (p+1)$, representing the binary expansion of the symbol over $\mathbb{F}_p$. The binary expansions are stacked on top of each other according to the given parity-check string. The interesting observation is that, from the sum of the binary strings over the reals, one can clearly obtain the binary expansion of the symbols in the sum, and then generate the residues modulo $p$ of the elements of the string to obtain the redundancy information needed for decoding. The obtained code is linear. 

Henceforth, we use the value $N$ to denote the length of the uniquely reconstructable strings \textit{\textbf{$h$-MC}} with added error-control redundancy. It is not to be confused with the parameter $N$ from Section~\ref{sec:lbmc} and Figure~\ref{fig:4} as the notation is reused to avoid clutter. Also, as before, we let $\cS_h(n)$ be a binary $B_h$ code constructed using the parity-check matrix of a binary code with minimum Hamming distance $\geq 2h +1$ and to add the syndrome redundancy, we use a BCH code of appropriate parameters. The main observation leading to the discussion of our scheme is that due to our encoding method which uses the complementation/bit flipping procedure, we require an unequal error-protection scheme. Recall that the substring $\br$ used in the construction described in Section III is the indicator vector for substring (of length $\sqrt{n}$ bits) flips. Errors in the $\br$ substring may clearly cause a burst of ``complementation errors'' due to the fact that $\br$ indicates if a string or its complement should be used. There are two approaches one can follow, by either encoding the string to handle a larger number of erasures independent on their location (the One-Step procedure) or by adding specialized redundancy to the $\br$ string (the Two-Step procedure).

\begin{figure} 
  \includegraphics[width=\linewidth]{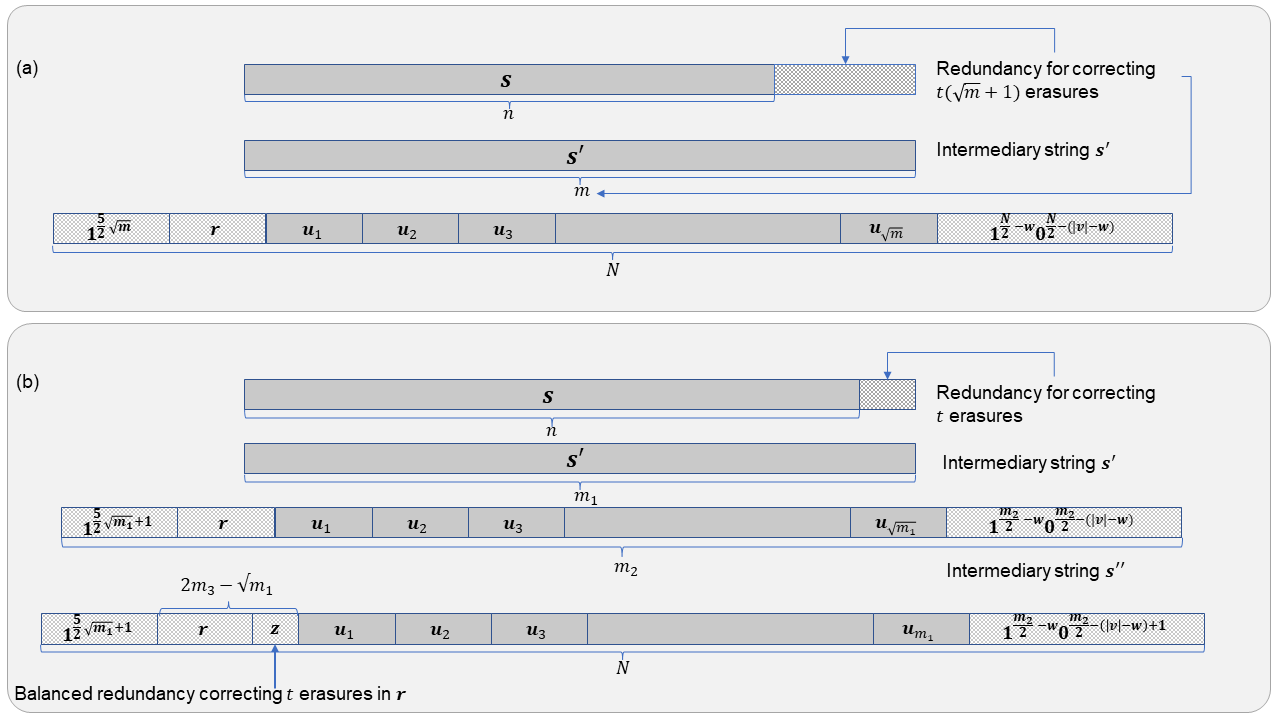}
  \caption{Illustration of the error-correction procedure. Subfigure (a) describes the One-Step error-correction procedure.The string is encoded to handle the worst-case erasure scenario. Subsequently, the approximate balancing construction from Section~\ref{sec:lbmc} is performed to generate a Dyck codestring. Note that $\bv = {\bf1}^{5/2 \sqrt{m}} \br \bu$ is as defined in Claim~\ref{cl:bal2}, and it pertains to the intermediary string $\bs'$. Here,  $w=\text{wt}(\bv)$. Subfigure (b) describes the Two-Step error-correction procedure. The string is encoded to be able to correct $t$ erasures. The approximate balancing procedure from Section~\ref{sec:lbmc} is performed to generate a Dyck string. In the next step, redundancy is added to the substring $\br$ that can be used to correct $t$ erasures. To preserve the Dyck property, ``balancing redundancy'' $\bz$ is appended to $\br,$ and additional bits are added to the prefix and suffix to obtain the desired codestring. Note that here $\bv = {\bf1}^{5/2 \sqrt{m_1}} \br \bu $ is as defined as in Claim~\ref{cl:bal2}, as applied to the intermediary string $\bs''$ depicted in Subfigure (b). Once again, we have $w = \text{wt}(\bv)$.} 
  \label{fig:encoding}
\end{figure}

The One-Step encoding method is illustrated in Figure~\ref{fig:encoding} and proceeds as follows:
\begin{itemize}
\item Each string $\bs \in \cS_h(n)$ is encoded using a BCH code into an intermediary string $\bs'$ of length $m,$ capable of correcting $t(\sqrt{m}+1)$ erasures. The redundancy required is at most $\lceil \frac{t}{2} \rceil (\sqrt{m}+1) \log (m+1).$ 
\item The intermediary string $\bs'$ of length $m$ is subsequently encoded via the balancing procedure described in Section~\ref{sec:lbmc}. The encoded balanced string has length $N$ and belongs to a \textit{\textbf{$h$-MC}} code capable of correcting up to $t$ composition erasures; here, $N=m+\frac{17}{2} \sqrt{m}$ which is at most 
$$n+ \lceil \frac{t}{2} \rceil   (\sqrt{m}+1) \log (m+1) + \frac{17}{2} \sqrt{m}.$$ 
The parameter $N$ can be further  bounded by above by 
$$  n+ \lceil \frac{t}{2} \rceil  (\sqrt{n} +1) \log n  + \frac{17}{2} \sqrt{n} + \epsilon_n \sqrt{n} \left( \lceil \frac{t}{2} \rceil  \log n + \frac{17}{2} \right) + \lceil \frac{t}{2} \rceil  \delta_n, $$ where $\epsilon_n  = \frac{ \lceil \frac{t}{2} \rceil  (\sqrt{m} + 1) \log (m+1)}{2n}$ and $\delta_n = \frac{ \lceil \frac{t}{2} \rceil  (\sqrt{m} +1) \log (m+1)}{ n}.$ 
\end{itemize}
As either the partial prefix-sum or the partial suffix-sum string has $\leq t (\sqrt{m}+1)$, the binary sum of the input strings can be recovered correctly. The decoding procedure for strings involved in the sum is identical to the one as described in Section~\ref{sec:lbmc}, and hence omitted.

\subsection{Two-Step Error-Correction}

As observed in the One-step scheme, errors in the substring $\br$ cause blocks of errors in the global string: Each erasure in $\br$ results in $\sqrt{m}$ additional erasures, where $m$ is the length of the approximately balanced strings. In order to overcome this issue in a more tailor-made manner, one can use unequal error-correction schemes that ensure that the binary sum of the $\br$ substring components across the input strings can be recovered independently from the rest of the string. The correctly reconstructed binary $\br$ sum can then be used for subsequent decoding of the complete collection of input strings. This Two-Step approach is illustrated in Figure~\ref{fig:encoding}.

As before, let $\cS_h(n)$ be a binary $B_h$ code constructed using the parity-check matrix of a code with minimum Hamming distance $\geq 2h +1$ (say, a BCH code). Furthermore, let $N$ denote the overall length of the \textit{\textbf{$h$-MC}} codestrings with added redundancy for mass error-correction. The encoding method is illustrated in Figure~\ref{fig:encoding}) and proceeds as follows:
\begin{itemize}
\item Each string $\bs \in \cS_h(n)$ is encoded into an intermediary strings $\bs' $ of length $m_1$ capable of correcting $t$ erasures, using a BCH code. The redundancy required is at most $t \log (m_1+1),$ and 
$$ m_1 \leq n + \lceil \frac{t}{2} \rceil \log (m_1 +1).$$
\item Each intermediary string $\bs'$ of length $m_1$ is encoded into a Dyck string using the procedure described in Section~\ref{sec:lbmc}, to arrive at a second intermediary string  $\bs"$ of length $m_2$, where 
$$m_2 = m_1 + \frac{17}{2} \sqrt{m_1}.$$
\item The substring $\br$ of the intermediary string $\bs"$ is encoded into a codestring $\br\br'$ of total length 
$m_3$, capable of correcting $t$ erasures. Let $m_4 = m_3 - \sqrt{m_1}$ denote the length of $\br'$. It is easy to see that $m_4 \leq \lceil \frac{t}{2} \rceil \log (m_3 +1).$ 
\item Since the string has to be balanced, $\br' =r'_1 r'_2 \dots r'_{m_4}$ is converted into $\bz = r'_1 \bar{r'}_1 r'_2 \bar{r'}_2\dots r'_{m_4-1} \bar{r'}_{m_4},$ where  $\bar{r'}_i = 1- r'_i.$
\item The balanced redundancy $\bz$ is appended to the $\br$ substring of the intermediary string $\bs".$ Also, a bit $1$ is added to the prefix of $\textbf{1}$s and a bit $0$ is appended to the suffix of $\textbf{0}$s to preserve the Dyck property of the string.
\end{itemize} 
The length of the coded string equals 
$$N  =  m_1 + \frac{17}{2} \sqrt{m_1} + 2(m_3 - \sqrt{m_1})  +2,$$
and upper-bounded in terms of the length $n$ as  
$$n+ \lceil \frac{t}{2} \rceil (\log n + \mu_n ) + \frac{17}{2} \sqrt{n}(1 + \nu_n) + \lceil \frac{t}{2} \rceil (\log n + \mu_n) + 2 \theta_n  + 2.$$ 
where $\mu_n  = \frac{\lceil \frac{t}{2} \rceil \log (m_1 + 1) + 1}{n},$ $\nu_n = \frac{\lceil \frac{t}{2} \rceil \log (m_1 +1 )}{2n},$ and $\theta_n = \frac{t \log (m_3 + 1) +1 }{\sqrt{n + \lceil \frac{t}{2} \rceil \log (m_1 +1)}}$.

\subsection{Composition Error-Correction Using String Integrals} \label{sec:direct}

One observation that is apparent from the previous examples is that erasures/errors caused in one mass propagate to one more error when used to reconstruct the real-valued sum of the strings. One simple means to mitigate this problem is to use running sums of symbols, in which case the errors cancel. A more precise explanation for how to perform encoding with this approach in mind is as follows. 

Without loss of generality, suppose that $t_p < t_s$. In this case, it is always possible for the errors in the suffix string to be such that we receive no additional information by considering both the prefix and suffix string, and so the problem at hand becomes to recover $\bs$ from a set of at most $n - t_p$ prefix compositions. 

\begin{claim} \label{cl:direct}
Suppose that $\cC(n,d) \subseteq \mathbb{F}_2^n$ is a code with minimum Hamming distance $d=\min \{ t_p, t_s \}+1.$ Let $\bs \in \mathbb{F}_2^n$ and fix $\emph{wt}(\bs) = w_0.$ Let $\tilde{\cM}_p(\bs)$ be the result of removing $t_p$ compositions from $\cM_p(\bs),$ and $t_s$ compositions from $\cM_s(\bs)$. Then, we can recover $\bs= s_1  s_2 \ldots s_n \in \{0,1\}^n$ from $\tilde{\cM}_p(\bs)$ provided that
$$ s_1 \, \, (s_1 + s_2)  \, \, (s_1 + s_2 + s_3) \, \, \ldots \, \, \sum_{j=1}^n s_j \in \cC(n,d).$$
\end{claim}
\begin{IEEEproof} Without loss of generality, assume that $t_p = \min \{ t_p, t_s \}$. The result follows since for $i \in [n]$ the value of the $i$-th component in the string $  s_1 \, \, (s_1 + s_2)  \, \, (s_1 + s_2 + s_3) \, \, \ldots \, \, \sum_{j=1}^n s_j  \in \cC(n,d)$ can be recovered by summing up the number of $1$s (modulo $2$) in the $i$-th prefix composition. The claim then follows since we know the lengths of the compositions that are missing from the set $\tilde{\cM}_p(\bs)$ and can hence recover the string $ s_1 \, \, (s_1 + s_2)  \, \, (s_1 + s_2 + s_3) \, \, \ldots \, \, \sum_{j=1}^n s_j,$ where $\bfs \in \mathbb{F}_2^n,$ from which $\bs$ can be then determined uniquely. 
Note that for the case that $t_s = \min \{ t_p, t_s \},$ since the weight of $\bs$ is known, a missing composition of a prefix of length $i$ can be recovered from the known composition of a suffix of length $n-i.$ Thus, $t_p+t_s$ missing compositions in $\tilde{\cM}_p(\bs)$ can be recovered from $\tilde{\cM}_s(\bs)$ and $w_0.$ This  concludes the proof.
\end{IEEEproof}

Using the result of Claim~\ref{cl:direct}, we can encode our mixture-strings using the following encoding technique:
\begin{itemize}
\item Given a string $\bs \in \{ 0,1\}^n,$ construct $I(\bs) = s_1 (s_1+s_2) (s_1 +s_2 + s_3) \dots (s_1+s_2 + \dots + s_n) = I(\bs)_1 I(\bs)_2  \dots I(\bs)_n$.
\item Encode $I(\bs)$ using a BCH code such that the resulting string $I(\bs)R'(\bs)$, where $R'(\bs)$ denotes the redundancy bits, is of length $m$ and can correct $\lfloor \frac{t}{2} \rfloor$ erasures. To ensure that the redundancy is itself properly balanced, we write $R(\br)_1 = R'(\br)_1 + I(\br)_1;$
$R(\br)_{2i} = \bar{R(\br)}_{2i} = 1 - R(\br)_{2i},$ and $R(\br)_{2i +1} = R'(\br)_{i} + R'(\br)_{i+1} + R(\br)_{2i},$ for all $i \in  [m-n-1],$ where $m,n$ are as described in the encoding scheme of Section~\ref{sec:direct}.
\item Encode $\bs$ as $\bs R(\bs)$.
\end{itemize}
To apply the above procedure, we need to be able to partition the prefix and suffix compositions of the $\bs R(\bs)$. This is easily achieved when $\bs R(\bs)$ is a substring of a Dyck string such that the composition of the prefix preceding the $\bs R(\bs)$-substring in the Dyck string is known. In particular, since the substring $\bs R(\bs)$ occurs after the runlength of $1$s in the construction of Section~\ref{sec:lbmc}, the prefix compositions of the constructed string $\bs R(\bs)$ can be recovered by subtracting the weight of the leading runlength of $1$s from the corresponding compositions. Consequently, $\bs R(\bs)$ satisfies the conditions of Claim~\ref{cl:direct} and the code is linear. Thus, the binary sum of multiple strings constructed using this technique also satisfies the conditions supporting Claim~\ref{cl:direct}.

In conclusion, the $t$-erasure-correcting Two-Step procedure can be replaced by the $\lfloor \frac{t}{2} \rfloor$ error-correcting technique outlined above. This results in an overall reduction of the redundancy by a factor of two, but adds to the encoding complexity.

\subsection{Correcting prefix-suffix mass substitution errors} \label{sec:substitution}

We now briefly turn our attention to describing how substitution errors in prefix-suffix compositions influence the reconstruction process. As already pointed out, in practice we often observe a larger mass being reported as a smaller mass due to losses of atoms during the fragmentation process. Such errors can be modeled as \emph{asymmetric errors} in which a bit equal to $1$ in a prefix or suffix   compositions is replaced by a $0$. For convenience, we refer to such errors as \textit{mass reducing substitution} errors. In the previous exposition, we illustrated the fact that the union of the prefix-suffix composition multisets and the sum of the input Dyck strings can be recovered even in the presence of a single erasure without additional redundancy. However, the same is not always true for mass reducing substitution errors, and hence the problem of correcting such errors is significantly more challenging. We illustrate this issue in Example~\ref{eg:mrsuberr}.   

\begin{example} \label{eg:mrsuberr}
\begin{enumerate}
\item Consider the following codebook of Dyck strings $\{ 110100,101010,110010,111000\},$ and select $\bt_1 = 111000, \bt_2 = 110100$. The error-free union of the prefix-suffix multisets of the strings equals 
$$\cM(\bt_1) \cup \cM(\bt_2) =\{1,1,1^2,1^2, 1^3, 01^2, 01^3, 01^3, 0^21^3, 0^21^3, 0^31^3, 0^31^3,  0^31^3, 0^31^3,  0^31^2, 0^31^2,0^31^1, 0^31^1, 0^3, 0^21, 0^2, 0^2, 0, 0 \}.$$ Using the defining property of Dyck strings, from $\cM(\bt_1) \cup \cM(\bt_2)$ one can reconstruct the multisets 
$$\cM_p(\bt_1) \cup \cM_p(\bt_2) = \{ 1,1,1^2,1^2, 1^3, 01^2, 01^3, 01^3, 0^21^3, 0^21^3, 0^31^3, 0^31^3 \}$$ 
and
$$\cM_s(\bt_1) \cup \cM_s(\bt_2) = \{0^31^3, 0^31^3,  0^31^2, 0^31^2,0^31^1, 0^31^1, 0^3, 0^21, 0^2, 0^2, 0, 0 \}.$$
Now, assume that one composition in the union, $1^2$, is erroneously read as $0^2$. Thus, in the presence of such an error, we obtain a different partition of compositions into prefix and suffix sets,
$$\cM_p(\bt_1) \cup \cM_p(\bt_2) = \{ 1,1,1^2,1^3, 01^2, 01^3, 01^3, 0^21^3, 0^21^3, 0^31^3, 0^31^3 \} ,$$ 
$$\cM_s(\bt_1) \cup \cM_s(\bt_2) = \{0^31^3, 0^31^3,  0^31^2, 0^31^2,0^31^1, 0^31^1, 0^3, 0^21, 0^2,0^2,  0^2, 0, 0 \}.$$ 
If it were known a priori that at most one mass reducing substitution error occurred, then an immediate conclusion would be that the composition of a prefix of length two was erroneously read as a suffix. Noting that the suffix compositions of the Dyck strings of length $n-2=4$ are $0^31^1, 0^31^1,$ it must be that the correct prefix compositions of length $2$ are $1^2$ and $1^2$. Thus, from the corrected multiset 
$$\cM_p(\bt_1) \cup \cM_p(\bt_2) = \{ 1,1,1^2,1^2,1^3, 01^2, 01^3, 01^3, 0^21^3, 0^21^3, 0^31^3, 0^31^3 \} ,$$ one can easily compute $\bt_1 + \bt_2 = 221100.$  

\item Next, consider the strings $\bt_2 = 110100$ and $\bt_3 = 110010.$ Their error-free union of prefix-suffix multisets is given by 
$$\cM(\bt_2) \cup \cM(\bt_3) =\{1,1,1^2,1^2, 01^2, 01^2, 01^3, 0^21^2, 0^21^3, 0^21^3, 0^31^3, 0^31^3,  0^31^3, 0^31^3,  0^31^2, 0^31^2,0^31^1, 0^31^1, 0^21, 0^21, 0^2, 01, 0, 0 \}.$$ 
Using once again the defining properties of Dyck strings, from $\cM(\bt_1) \cup \cM(\bt_2)$ we can reconstruct the multisets 
$$\cM_p(\bt_2) \cup \cM_p(\bt_3) = \{ 1,1,1^2,1^2, 01^2, 01^2, 01^3, 0^21^2, 0^21^3, 0^21^3, 0^31^3, 0^31^3 \}$$
and 
$$\cM_s(\bt_2) \cup \cM_s(\bt_3) = \{ 0^31^3, 0^31^3,  0^31^2, 0^31^2,0^31^1, 0^31^1, 0^21, 0^21, 0^2, 01, 0, 0  \}.$$ 
In addition, let us also examine the strings $\bt_1 = 111000$ and $\bt_3 = 110010.$ Their error-free union of prefix-suffix multiset is given by 
$$\cM(\bt_1) \cup \cM(\bt_3) =\{1,1,1^2,1^2,1^3, 01^2, 01^3, 0^21^2, 0^21^3, 0^21^3, 0^31^3, 0^31^3,  0^31^3, 0^31^3,  0^31^2, 0^31^2, 0^31^1, 0^31^1, 0^3, 0^21, 0^2, 01, 0, 0 \}.$$ 
In this case,
$$\cM_p(\bt_2) \cup \cM_p(\bt_3) = \{ 1,1,1^2,1^2,1^3, 01^2, 01^3, 0^21^2, 0^21^3, 0^21^3, 0^31^3, 0^31^3 \} ,$$ 
$$\cM_s(\bt_1) \cup \cM_s(\bt_3) = \{ 0^31^3, 0^31^3,  0^31^2, 0^31^2, 0^31^1, 0^31^1, 0^3, 0^21, 0^2, 01, 0, 0 \}.$$ 
Given the erroneous union of prefix-suffix composition multiset of two strings 
$$\{1,1,1^2,1^2, 0^3, 01^2, 01^3, 0^21^2, 0^21^3, 0^21^3, 0^31^3, 0^31^3,  0^31^3, 0^31^3,  0^31^2, 0^31^2,0^31^1, 0^31^1, 0^21, 0^21, 0^2, 01, 0, 0 \},$$  the following scenarios are possible: \\
a) In $\cM(\bt_2) \cup \cM(\bt_3),$ the prefix composition $01^2$ was changed to $0^3$, or \\
b) In $ \cM(\bt_1) \cup \cM(\bt_3),$ the prefix composition $1^3$ was changed to $0^21.$
Thus, in the presence of even a single mass reducing composition error, without additional coding redundancy, neither the union of the prefix-suffix composition multiset nor the sum of the strings can be uniquely reconstructed.

\item In the final example, let us revisit the strings $\bt_2 = 110100$ and $\bt_3 = 110010.$ 
Given the erroneous composition multiset 
$$\tilde{\cM}(\bt_2) \cup \tilde{\cM}(\bt_3) =\{1,1,1^2, 01, 01^2, 01^2, 01^3, 0^21^2, 0^21^3, 0^21^3, 0^31^3, 0^31^3,  0^31^3, 0^31^3,  0^31^2, 0^31^2,0^31^1, 0^31^1, 0^21, 0^21, 0^2, 01, 0, 0 \}, $$ 
one can recover
$$\tilde{\cM}_p(\bt_2) \cup \tilde{\cM}_p(\bt_3) = \{ 1,1,1^2, 01, 01^2, 01^2, 01^3, 0^21^2, 0^21^3, 0^21^3, 0^31^3, 0^31^3 \}$$
and 
$$\tilde{\cM}_s(\bt_2) \cup \tilde{\cM}_s(\bt_3) = \{ 0^31^3, 0^31^3,  0^31^2, 0^31^2,0^31^1, 0^31^1, 0^21, 0^21, 0^2, 01, 0, 0  \}.$$ 
Noting that the prefix compositions $1^2, 01$ are not compatible with the suffix compositions $0^31^1, 0^31^1,$ we conclude that one of these four compositions must be in error. Hence, either the prefix sum string $211110$ or the suffix sum string $220110$ is correct and we can examine both settings to determine the possible solutions.
\end{enumerate}
\end{example}
In summary, the examples above illustrate that a single mass reducing error can always be detected but not necessarily corrected. For the special case where a composition of a prefix or suffix of length $i$ is replaced by the composition of a prefix or suffix of the same length, a simple modification to the coding technique described for correcting $t$ mass erasures can be used to correct $t$ mass substitution errors. To this end, we make the following two observations.

Although based on the pigeon-hole principle it is true that in the either the union of the prefix composition multisets or the union of the suffix composition multisets is such that it contains at most $\lfloor \frac{t}{2} \rfloor$ mass reducing errors, it is not possible to identify which of the two multisets contains fewer errors. 
As a result, the Two-Step encoding procedure has to involve a \textit{\textbf{$h$-MC}} code capable of correcting $t$ substitution errors. This results in a four times higher redundancy compared to that used for correcting missing mass errors. 

\section{Open Problems}

Many combinatorial and coding-theoretic problems related to string reconstruction from prefix-suffix compositions remain open. A sampling is listed below.
\begin{itemize}
\item Our techniques for converting a binary string of length $n$ into strings that are both Dyck and belong to a $B_h$ codebook have suboptimal redundancy. We seek methods that can reduce our overhead and at the same time, offer low encoding and decoding complexity.
\item In practice, one often encounters nonbinary alphabets, as polymers can be synthesized to have highly different masses and chemical properties. The question remains to generalize our approach for nonbinary alphabets. Furthermore, it is of interest to investigate such coding techniques for strings that have some form of balanced symbol contents or masses confined to a certain interval.
\item It remains an open question to characterize all the missing mass errors that can be corrected by simply utilizing the Dyck, $B_h$ properties of strings and the presence of both prefix and suffix masses. 
\item At this point, we have no efficient means for correcting mass reducing (or, mass increasing) substitution errors in our mixtures. A solution to this problem can have interesting and important implications in the field of polymer-based data storage.
\end{itemize}

\section*{Acknowledgment}
The work was funded by the DARPA Molecular Informatics Program, the NSF portion of funding from the SemiSynBio program and the NSF grants CIF 2008125 and 1618366. Parts of the work were presented at 
the Information Theory Workshop at Riva la Guarda, Italy, 2020 (moved to 2021).
\ifCLASSOPTIONcaptionsoff
  \newpage
\fi

\bibliography{IEEEabrv,biblio}
\bibliographystyle{IEEEtran}
\end{document}